\makeatletter
\let\@twosidetrue\@twosidefalse
\let\@mparswitchtrue\@mparswitchfalse
\makeatother

\documentclass{llncs}
\usepackage[T1]{fontenc}
\usepackage[latin2]{inputenc}
\usepackage[english]{babel}
\usepackage{dsfont, complexity, lmodern, lipsum} 
\usepackage{amsmath}
\usepackage{amssymb}
\usepackage{float}
\usepackage{graphicx}
\usepackage{tikz}
\usepackage{xcolor,xspace}
\usepackage{paralist}
\usepackage{refcount}
\usepackage[algoruled,vlined]{algorithm2e}

\usepackage{thmtools, mathtools}
\usepackage{thm-restate}

\makeatletter 
\def\@seccntformat#1{\@ifundefined{#1@cntformat}%
   {\csname the#1\endcsname\quad}  
   {\csname #1@cntformat\endcsname}
}
\let\oldappendix\appendix 
\renewcommand\appendix{%
    \oldappendix
    \newcommand{\section@cntformat}{\appendixname~\thesection\quad}
}
\makeatother

\usepackage[textsize=tiny,textwidth=2.2cm,shadow]{todonotes}

\usepackage{hyperref}

\DeclareMathOperator{\rank}{r}
\DeclareMathOperator{\propose}{\pi}
\DeclareMathOperator{\refuse}{\rho}
\DeclareMathOperator{\argmin}{argmin}
\DeclareMathOperator{\argmax}{argmax}

\definecolor{MyPurple}{RGB}{197,0,205}
\usetikzlibrary{calc, lindenmayersystems, decorations, decorations.pathmorphing, decorations.markings, shapes, shapes.geometric, arrows, backgrounds, patterns, decorations.pathreplacing, positioning}
\pgfdeclarelayer{background}
\pgfdeclarelayer{foreground}
\pgfsetlayers{background,main,foreground}

\tikzstyle{vertex} = [circle, draw=black, fill=black, scale= 0.5]
\tikzstyle{edgelabel} = [rectangle, fill=white]
\tikzstyle{arrow} = [line width=0.8mm,-implies,double, double distance=0.8mm]
\tikzstyle{dashedpointedline} = [line width=0.2mm,dashed,dash pattern=on 2mm off 1mm,
decoration={markings,
	mark=at position 1 with {\arrow[line width=0.2mm, scale= 1.8]{>}}
},
postaction={decorate}
] 
\tikzstyle{pointedline} = [line width=0.3mm,
decoration={markings,
	mark=at position 1 with {\arrow[line width=0.2mm, scale= 2]{>}}
},
postaction={decorate}
] 
\tikzstyle{reversepointedline} = [line width=0.3mm,
decoration={markings,
	mark=at position 0.02 with {\arrow[line width=0.2mm, scale= 2]{<}}
},
postaction={decorate}
] 

\tikzstyle{pointedline2} = [line width=0.3mm,
decoration={markings,
	mark=at position 0.65 with {\arrow[line width=0.2mm, scale= 2]{>}}
},
postaction={decorate}
] 

\spnewtheorem{obs}{Observation}{\bfseries}{\itshape}
\spnewtheorem{pr}{Problem}{\bfseries}{\rmfamily}
\spnewtheorem{ex}{Example}{\bfseries}{\rmfamily}
\newcommand{\true}{\mathsf{true}}
\newcommand{\false}{\mathsf{false}}
\newcommand{\myqed}{$\blacksquare$}
\newcommand{\myproof}{\noindent\textit{Proof. }}

\newcommand{\inp}{\textsf{Input: }} 
 
\newcommand{\ques}{\textsf{Question: }} 

\newcommand{\st}{\,:\,}

\begin{document}
\title{New and simple algorithms for stable flow problems}
\author{
\'{A}gnes Cseh\inst{1}\thanks{Supported by Cooperation of Excellences Grant (KEP-6/2018), by the Ministry of Human Resources under its New National Excellence Programme (UNKP-18-4-BME-331), the Hungarian Academy of Sciences under its Momentum Programme (LP2016-3/2016), its J\'anos Bolyai Research Fellowship, and OTKA grant K128611.} \and 
Jannik Matuschke\inst{2}\thanks{Partially supported by COST Action IC1205 on Computational Social Choice.}}

\institute{Institute of Economics, Hungarian Academy of Sciences and\\ Corvinus University of Budapest, e-mail: \texttt{cseh.agnes@krtk.mta.hu} \\ \and
TUM School of Management, Technische Universit\"at M\"unchen, e-mail: \texttt{jannik.matuschke@tum.de}}

\date{}
\maketitle

\begin{abstract}
Stable flows generalize the well-known concept of stable matchings to markets in which transactions may involve several agents, forwarding flow from one to another. An instance of the problem consists of a capacitated directed network in which vertices express their preferences over their incident edges. A network flow is stable if there is no group of vertices that all could benefit from rerouting the flow along a walk.

Fleiner~\cite{Fle14} established that a stable flow always exists by reducing it to the stable allocation problem.
We present an augmenting path algorithm for computing a stable flow, the first algorithm that achieves polynomial running time for this problem without using stable allocations as a black-box subroutine. We further consider the problem of finding a stable flow such that the flow value on every edge is within a given interval. For this problem, we present an elegant graph transformation and based on this, we devise a simple and fast algorithm, which also can be used to find a solution to the stable marriage problem with forced and forbidden edges.

Finally, we study the stable multicommodity flow model introduced by Kir\'{a}ly and Pap~\cite{KP13a}. The original model is highly involved and allows for commodity-dependent preference lists at the vertices and commodity-specific edge capacities. We present several graph-based reductions that show equivalence to a significantly simpler model. We further show that it is $\NP$-complete to decide whether an integral solution exists.

\end{abstract}

\section{Introduction}
\label{sec:st_fl}

Stability is a well-known concept used for matching markets without monetary transactions~\cite{Rot84}.
A stable solution provides certainty that no two agents are willing to selfishly modify the market situation. Stable matchings were first formally defined in the seminal paper of Gale and Shapley~\cite{GS62}. They described an instance of the college admission problem and introduced the terminology based on marriage that since then became wide-spread. 
Besides this initial application, variants of the stable matching problem are widely used in employer allocation markets~\cite{RS90}, university admission decisions~\cite{BS99,BDK10}, campus housing assignments~\cite{CS02,PPR08} and bandwidth allocation~\cite{GLMMRV07}. A recent honor proves the currentness and importance of results in the topic: in 2012, Lloyd S. Shapley and Alvin E. Roth were awarded the Sveriges Riksbank Prize in Economic Sciences in Memory of Alfred Nobel for their outstanding results on market design and matching theory. 

In the classic stable marriage problem, we are given a bipartite graph, where the two classes of vertices represent men and women, respectively. Each vertex has a strictly ordered preference list over his or her possible partners. A matching is \emph{stable} if it is not \emph{blocked} by any edge, that is, no man-woman pair exists who are mutually inclined to abandon their partners and marry each other~\cite{GS62}.

In practice, the stable matching problem is mostly used in one of its capacitated variants, which are the stable many-to-one matching, many-to-many matching and allocation problems. The \emph{stable flow} problem can be seen as a high-level generalization of all these settings. 
As the most complex graph-theoretical generalization of the stable marriage model, it plays a crucial role in the theoretical understanding of the power and limitations of the stability concept. From a practical point of view, stable flows can be used to model markets in which interactions between agents can involve chains of participants, e.g., supply chain networks involving multiple independent companies.

In the stable flow problem, a directed network with preferences models a market situation. Vertices are vendors dealing with some goods, while edges connecting them represent possible deals. Through his preference list, each vendor specifies how desirable a trade would be to him. Sources and sinks model suppliers and end-consumers. A feasible network flow is stable, if there is no set of vendors who mutually agree to modify the flow in the same manner. A blocking walk represents a set of vendors and a set of possible deals so that all of these vendors would benefit from rerouting some flow along the blocking walk.

\paragraph{Literature review.} The notion of stability was extended to so-called ``vertical networks'' by Ostrovsky in 2008~\cite{Ost08}. Even though the author proves the existence of a stable solution and presents an extension of the Gale-Shapley algorithm, his model is restricted to unit-capacity acyclic graphs. Stable flows in the more general setting were defined by Fleiner~\cite{Fle14}, who reduced the stable flow problem to the stable allocation problem. Since then, the stable flow problem has been investigated in several papers~\cite{FJJT18,FJST18,Jag17,LN17}.
Recently, stable flows have been used to derive conflict-free routings in multi-layer graphs~\cite{SW15}.

The best currently known computation time for finding a stable flow is $\mathcal{O}(|E| \log |V|)$ in a network with vertex set $V$ and edge set~$E$. This bound is due to Fleiner's reduction to the stable allocation problem and its fastest solution described by Dean and Munshi~\cite{DM10}. Since the reduction takes $\mathcal{O}(|V|)$ time, it does not change the instance size significantly, and the weighted stable allocation problem can be solved in $\mathcal{O}(|E|^2 \log |V|)$ time~\cite{DM10}, the same holds for the maximum weight stable flow problem. The Gale-Shapley algorithm can also be extended for stable flows~\cite{CMS13}, but its straightforward implementation requires pseudo-polynomial running time, just like in the stable allocation problem.

It is sometimes desirable to compute stable solutions using certain forced edges or avoiding a set of forbidden edges.
This setting has been an actively researched topic for decades~\cite{CM16,DFFS03,FIM07,GI89,Knu76}. This problem is known to be solvable in polynomial time in the one-to-one matching case, even in non-bipartite graphs~\cite{FIM07}. Though Knuth presented a combinatorial method that finds a stable matching in a bipartite graph with a given set of forced edges or reports that none exists~\cite{Knu76}, all known methods for finding a stable matching with both forced and forbidden edges exploit a somewhat involved machinery, such as rotations~\cite{GI89}, LP techniques~\cite{Fed92,Fed94,ILG87} or reduction to other advanced problems in stability~\cite{DFFS03,FIM07}.

In many flow-based applications, various goods are exchanged. Such problems are usually modeled by multicommodity flows~\cite{Jew66}. A maximum multicommodity flow can be computed in strongly polynomial time~\cite{Tar86}, but even when capacities are integer, all optimal solutions might be fractional, and finding a maximum integer multicommodity flow is $\NP$-hard~\cite{GJ79}. 
Kir\'aly and Pap~\cite{KP13a} introduced the concept of stable multicommodity flows, in which edges have preferences over which commodities they like to transport and the preference lists at the vertices may depend on the commodity. They show that a stable solution always exists, but it is $\PPAD$-hard to find one.

\paragraph{Our contribution and structure.} In this paper we discuss new and simplified algorithms and complexity results for three differently complex variants of the stable flow problem. Section~\ref{sec:prel_sf} contains preliminaries on stable flows.
\begin{itemize}
	\item In Section~\ref{polalg_sf} we present a \textit{polynomial algorithm for stable flows}. To derive an efficient solution method operating directly on the flow network, we combine the well-known pseudo-polynomial Gale-Shapley algorithm and the proposal-refusal pointer machinery known from stable allocations into an augmenting path algorithm for computing a stable flow. Besides polynomial running time, the method has the advantage that it is easy to implement and that it provides new insights into the structure of the stable flow problem, which we exploit in later sections.
    \item Then, in Section~\ref{sec:re_sf} \textit{stable flows with restricted intervals} are discussed. We provide a simple combinatorial algorithm to find a flow with flow value within a pre-given interval for each edge. Surprisingly, our algorithm directly translates into a very simple new algorithm for the problem of stable matchings with forced and forbidden edges in the classical stable marriage case. Unlike the previously known methods, our result relies solely on elementary graph transformations. 
    \item Finally, in Section~\ref{sec:smcf} we study \textit{stable multicommodity flows}. First, we answer an open question posed in~\cite{KP13a} by providing tools to simplify stable multicommodity flow instances to a great extent. In particular, we show that it is without loss of generality to assume that no commodity-specific preferences at the vertices and no commodity-specific capacities on the edges exist. Then, we reduce \textsc{3-sat} to the integral stable multicommodity flow problem and show that it is $\NP$-complete to decide whether an integral solution exists even if the network in the input has integral capacities only.
\end{itemize}

\section{Preliminaries}
\label{sec:prel_sf}

A \emph{network} $(D, c)$ consists of a directed graph $D = (V, E)$ and a capacity function $c:E\rightarrow\mathbb{R}_{\geq 0}$ on its edges. The vertex set of $D$  has two distinct elements, also called \emph{terminal vertices}: a source $s$, which has outgoing edges only and a sink $t$, which has incoming edges only. Besides differentiating between the source and the sink, we will assume that $D$ does not contain loops or parallel edges, and every vertex $v \in V \setminus \{s, t\}$ has both incoming and outgoing edges. These three assumptions are without loss of generality and only for notational convenience.
We denote the set of edges leaving a vertex $v$ by  $\delta^{+}(v)$ and  the set of edges running to $v$ by  $\delta^{-}(v)$.

\begin{definition}[flow]
	Function $f:E\rightarrow\mathbb{R}_{\geq 0}$ is a \emph{flow} if it fulfills both of the following requirements:
	\begin{enumerate}
		\item capacity constraints: $f(uv) \leq c(uv)$ for every $uv \in E$;
		\item flow conservation: $\sum_{uv \in E}{f(uv)} = \sum_{vw \in E}{f(vw)}$ for all $v \in V \setminus \{s, t\}$.
	\end{enumerate}
\end{definition}

A stable flow instance is a triple $\mathcal{I} = (D, c, r)$. It comprises a network $(D,c)$ and $r$, a ranking function that induces for each vertex an ordering of their incident edges. 
Each non-terminal vertex ranks its incoming and also its outgoing edges strictly and separately. 
Formally, $r = (r_v)_{v \in V \setminus \{s,t\}}$, contains an injective function $r_v : \delta^+(v) \cup \delta^-(v) \rightarrow \mathbb{R}$ for each $v \in V \setminus \{s,t\}$.
We say that $v$ \emph{prefers} edge $e$ to~$e'$ if $\rank_v(e) < \rank_v(e')$. Terminals do not rank their edges, because their preferences are irrelevant with respect to the following definition.

\begin{definition}[blocking walk, stable flow]
\label{def:sf}
	A \emph{blocking walk} of flow $f$ is a directed walk $W=\langle v_{1}, v_{2}, ..., v_{k} \rangle$ such that all of the following properties hold:
	\begin{enumerate}
		\item $f(v_iv_{i+1}) < c(v_iv_{i+1})$, for each edge $v_iv_{i+1}$, $i=1, ...,k-1$;
		\item $v_{1} = s$ or there is an edge $v_{1}u$ such that $f(v_{1}u)>0$ and $\rank_{v_1} (v_{1} v_{2}) < \rank_{v_1} (v_{1}u)$; 
		\item $v_{k} = t$ or there is an edge $wv_{k}$ such that $f( wv_{k})>0$ and $\rank_{v_k} (v_{k-1} v_{k}) < \rank_{v_k} (wv_{k})$.
	\end{enumerate}
	A flow is \emph{stable}, if there is no blocking walk with respect to it in the graph.
\end{definition}

Intuitively, a blocking walk is an unsaturated walk in the graph so that both its starting vertex and its end vertex are inclined to reroute some flow along it. Notice that the preferences of the internal vertices of the walk do not matter in this definition.

Unsaturated walks fulfilling point~2 are said to \emph{dominate} $f$ at start, while walks fulfilling point~3 dominate $f$ at the end. We can say that a walk blocks $f$ if it dominates $f$ at both ends.


\begin{problem} \textsc{sf}\ \\
	\inp $\mathcal{I} = (D, c, r)$; a directed network $(D,c)$ and $r$, the preference ordering of vertices. \\
	\ques Is there a stable flow $f$?
\end{problem}

\begin{theorem}[Fleiner~\cite{Fle14}]
\label{same_flow_value}
	\textsc{sf} always has a stable solution and it can be found in polynomial time. Moreover, for a fixed \textsc{sf} instance, each edge incident to $s$ or $t$ has the same value in every stable flow.
\end{theorem} 

This result is based on a reduction to the stable allocation problem. The second half of Theorem~\ref{same_flow_value} can be seen as the flow generalization of the so-called \emph{Rural Hospitals Theorem} known for stable matching instances~\cite{GS85}. While Theorem~\ref{same_flow_value} implies that all stable flows have equal value, we remark that this value can be much smaller than that of a maximum flow in the network. In Example~\ref{ex:smallsf} we demonstrate a gap of $\Omega(|E|)$.

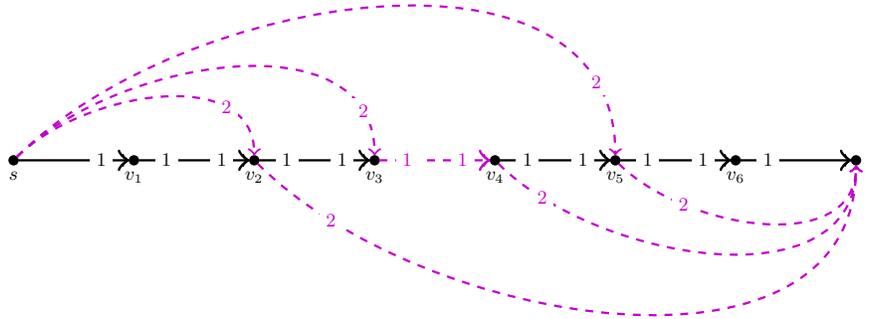
\begin{figure}[t]
	\centering
		\begin{tikzpicture}[scale=0.8, transform shape,->]
		\pgfmathsetmacro{\b}{2}
		\pgfmathsetmacro{\d}{2}
		\pgfmathsetmacro{\a}{90}
		\pgfmathsetmacro{\c}{45}
		
		\node[vertex, fill = black, label=below:$s$] (s) at (0, 0) {};
		\node[vertex, label=below:$v_1$] (v1) at (\b, 0) {};
		\node[vertex, label=below:$v_2$] (v2) at (2*\b, 0) {};
		\node[vertex, label=below:$v_3$] (v3) at (3*\b, 0) {};
		\node[vertex, label=below:$v_4$] (v4) at (4*\b, 0) {};
		\node[vertex, label=below:$v_5$] (v5) at (5*\b, 0) {};
		\node[vertex, label=below:$v_6$] (v6) at (6*\b, 0) {};
		\node[vertex, fill = black, label={[xshift=3mm, yshift=-7mm]$t$}] (t) at (7*\b, 0) {};
				
		\draw [pointedline] (s) -- node[edgelabel, near end] {1} (v1);
		\draw [pointedline] (v1) -- node[edgelabel, near start] {1} node[edgelabel, near end] {1} (v2);
		\draw [pointedline] (v2) --  node[edgelabel, near start] {1} node[edgelabel, near end] {1} (v3);
		\draw [pointedline, MyPurple, dashed] (v3) --  node[edgelabel, near start] {1} node[edgelabel, near end] {1} (v4);
		\draw [pointedline] (v4) --  node[edgelabel, near start] {1} node[edgelabel, near end] {1} (v5);
		\draw [pointedline] (v5) --  node[edgelabel, near start] {1} node[edgelabel, near end] {1} (v6);
		\draw [pointedline] (v6) --  node[edgelabel, near start] {1} (t);
		
		\draw [pointedline, MyPurple, dashed] (s) to[out=\c,in=\a] node[edgelabel,very  near end] {2} (v5);		
		\draw [pointedline, MyPurple, dashed] (s) to[out=\c,in=\a]  node[edgelabel,very  near end] {2}(v3);
		\draw [pointedline, MyPurple, dashed] (s) to[out=\c,in=\a]  node[edgelabel, near end] {2} (v2);
		
		\draw [pointedline, MyPurple, dashed] (v2) to[out=-\c,in=-\a] node[edgelabel, very near start] {2} (t);
		\draw [pointedline, MyPurple, dashed] (v4) to[out=-\c,in=-\a] node[edgelabel, very near start] {2} (t);
		\draw [pointedline, MyPurple, dashed] (v5) to[out=-\c,in=-\a] node[edgelabel,  near start] {2} (t);
		\end{tikzpicture}
\caption{The edge labels indicate the ranking of each edge at a vertex. For example, $v_3$prefers receiving flow from $v_2$ to receiving flow from~$s$. The maximum flow (marked by dashed colored edges) has value~3 in this unit-capacity network, while the unique stable flow is of value~1 and is sent along the path $\langle s, v_1, v_2, ..., t \rangle$. It is easy to see that this instance can be extended to demonstrate the ratio~$\Omega(|E|)$.}
\label{fi:maximum_maximal}
\end{figure}

\begin{ex}[Small stable flow value]
\label{ex:smallsf}
Flows with no unsaturated terminal-terminal paths are \emph{maximal} flows. We know that every stable flow is maximal and it is folklore that the ratio of the size of maximal and maximum flows can be of~$\mathcal{O}(|E|)$. As the instance in Fig.~\ref{fi:maximum_maximal} demonstrates, this ratio can also be achieved by the size of a stable flow vs.~that of a maximum flow.
\end{ex}

\section{A polynomial-time augmenting path algorithm for stable flows}
\label{polalg_sf}

Using Fleiner's construction~\cite{Fle14}, a stable flow can be found efficiently by computing a stable allocation in a transformed instance instead. Another approach is adapting the widely used Gale-Shapley algorithm to \textsc{sf}. As described in~\cite{CMS13}, this yields a preflow-push type algorithm, in which vertices forward or reject excessive flow according to their preference lists. While this algorithm has the advantage of operating directly on the network without transformation to stable allocation, its running time is only pseudo-polynomial.

In the following, we describe a polynomial time algorithm to produce a stable flow that operates directly on the network $D$. Our method is based on the well-known augmenting path algorithm of Ford and Fulkerson~\cite{FF62}, also used by Ba\"iou and Balinski~\cite{BB00} and Dean and Munshi~\cite{DM10} for stability problems. The main idea is to introduce proposal and refusal pointers to keep track of possible Gale-Shapley steps and execute them in bulk. Each such iteration corresponds to augmenting flow along an $s$-$t$-path or a cycle in a restricted residual network.

\subsection{Our algorithm}
In the algorithm, every vertex (except for the sink) is associated with two pointers, the \emph{proposal pointer} and the \emph{refusal pointer}. 
Throughout the course of the algorithm, the proposal pointer traverses the outgoing edges of the vertex in order of decreasing preference while the refusal pointer traverses its incoming edges in order of increasing preference. For the source $s$, we assume an arbitrary preference order. 
Starting with the $0$-flow, the algorithm iteratively augments the flow along a path or cycle in the graph induced by the pointers. This graph consists of the edges pointed at by the proposal pointers and the reversals of the edges pointed at by the refusal pointer.

After each augmentation step, pointers pointing at saturated or refused edges are advanced.
The algorithm terminates when the proposal pointer of the source has traversed all its outgoing edges. We prove that when this happens, the algorithm has found a stable flow. 
As in each iteration, at least one pointer is advanced, the running time of the algorithm is polynomial in the size of the graph. The complete algorithm is listed as Algorithm~\ref{alg:poly_sf}. In the following we describe the individual parts in detail.

\paragraph{Initializing and updating pointers.}
For notational convenience, we introduce two artificial elements, $*$ at the top and $\emptyset$ at the bottom of each preference list with the convention $\rank_v(*) = -\infty$ and $\rank_v(\emptyset) = \infty$.

Every vertex $v \in V \setminus \{t\}$ is associated with a \emph{proposal pointer} $\propose[v]$ and a \emph{refusal pointer} $\refuse[v]$, both pointing to elements on the preference list. 
Initially, $\propose[v]$ points to the most preferred outgoing edge on $v$'s preference list, i.e., the entry right after $*$, whereas $\refuse[v]$ is inactive, which is denoted by $\refuse[v] = \emptyset$. 
We also set $\refuse[t] = \emptyset$ for notational convenience (we will never change $\refuse[t]$ during the algorithm). Note that this implies $\rank_v(\refuse[t]) = \infty$.

The pointers at $v$ are advanced through the procedure \textsc{AdvancePointers}$(v)$; see Algorithm~\ref{alg:poly_sf}, lines~\ref{alg-line:update-pointers-start}-\ref{alg-line:update-pointers-end} for a formal listing. A call of this procedure works as follows:
 \begin{itemize}
 \item If $\propose[v]$ is active, it is advanced to point to the next less-preferred outgoing edge on $v$'s preference list (lines \ref{alg-line:proposal-advance-start}-\ref{alg-line:proposal-advance-end}). If all of $v$'s outgoing edges have been traversed, $\propose[v]$ reaches its inactive state, i.e., $\propose[v] = \emptyset$, and $\refuse[v]$ gets advanced from its inactive state to pointing to the least-preferred incoming edge on $v$'s preference list. Note that in this latter case, the state of $\propose[v]$ changes from active to inactive between line~\ref{alg-line:proposal-advance-start} and line~\ref{alg-line:refusal-pointer-advance-start}, and thus both if-conditions are fulfilled in the same call of the procedure.
 \item If $\propose[v]$ is already inactive, the refusal pointer $\refuse[v]$ gets advanced to the next more-preferred incoming edge on the preference list (lines~\ref{alg-line:refusal-pointer-advance-start}-\ref{alg-line:refusal-pointer-advance-end}). Once $\refuse[v]$ traversed $v$'s most preferred incoming edge, we set $\refuse[v] = *$, denoting all incoming edges of $v$ have been refused (the procedure will not be called again for this vertex after this point).
\end{itemize}

\begin{algorithm}[t]
	\SetKwProg{myproc}{procedure}{}{}
	\SetKwFunction{AdvancePointers}{AdvancePointers}
\caption{Augmenting path algorithm for stable flows}
\label{alg:poly_sf}

  \tcp{Initialize proposal pointers to point at most-preferred outgoing edges, refusal pointers inactive.}
  Set $\propose[v] := \argmin_{vw \in E} \rank_v(vw)$ and $\refuse[v] := \emptyset$ for all $v \in V$.\\
  Set $f := 0$.\\[0.2cm]
  
  \tcp{Ensure pointers only point to residual, non-refused edges.}
  \While{$\exists\, uv \in E_{H_{\propose, \refuse}}$ with $c_f(uv) = 0$ or $(\propose[u] = uv$ and $\rank_v(uv) \geq \rank_v(\refuse[v]))$\label{alg-line:pointer-update-start}}{
    \AdvancePointers($u$)\label{alg-line:pointer-update-end}
 }
  \vspace{0.2cm}
  
  \tcp{Stop once proposal pointer of source becomes inactive.}
  \If{$\propose[s] = \emptyset$\label{alg-line:termination-start}}{
    \Return $f$\label{alg-line:termination-end}
  }
  \vspace{0.2cm}
  
  \tcp{Augment flow along path/cycle induced by proposal and refusal pointers.}
  Let $W$ be an $s$-$t$-path or cycle in $H_{\propose, \refuse}$. 
  \label{alg-line:augmentation-start}\\
  Set $\Delta := \min_{e \in W} c_f(e)$. \\
	Augment $f$ by $\Delta$ along $W$. 
	\label{alg-line:augmentation-end}\\[0.2cm]
	
	\tcp{Repeat.}
	Goto line 3.\label{alg-line:repeat}
	\\[0.3cm]

	\SetKwProg{myproc}{procedure}{}{}
	\SetKwFunction{AdvancePointers}{AdvancePointers}
\myproc{\AdvancePointers \textup{($v$)}\label{alg-line:update-pointers-start}}{%

  \tcp{If proposal pointer is active, advance it to next less-preferred outgoing edge.}
  
  \If{$\propose[v] \neq \emptyset$%
      \label{alg-line:proposal-advance-start}}{
      
    Set $P := \{vw \in E : \rank_v(vw) > \rank_v(\propose[v])\} \cup \{\emptyset\}$.
      \label{alg-line:advance-propose-set}\\ 
    
    Set $\propose[v] := \argmin_{e \in P} \rank_v(e)$.
      \label{alg-line:advance-propose}\label{alg-line:proposal-advance-end}\\
  }
  \vspace{0.2cm}
  
  \tcp{If proposal pointer has passed all edges, advance refusal pointer to next more-preferred incoming edge.}
  
  \If{$\propose[v] = \emptyset$ and $\refuse[v] \neq *$%
      \label{alg-line:refusal-pointer-advance-start}}{
    Set $R := \{uv \in E : \rank_v(uv) < \rank_v(\refuse[v])\} \cup \{*\}$.
      \label{alg-line:advance-refuse-set}\\
    Set $\refuse[v] := \argmax_{e \in R} \rank_v(e)$.
      \label{alg-line:refusal-pointer-advance}\label{alg-line:refusal-pointer-advance-end}\label{alg-line:update-pointers-end}\\
  }
}
\end{algorithm}

\paragraph{The helper graph.}
With any state of the pointers $\propose, \refuse$, we associate a helper graph $H_{\propose, \refuse}$. It has the same vertex set as $D$ and the following edge set:
\begin{align*}
E_{H_{\propose, \refuse}} := & \ \{\propose[v] \,:\, v \in V \setminus \{t\},\, \propose[v] \neq \emptyset\} \\
& \qquad \cup \, \{\operatorname{rev}(\refuse[v]) \,:\, v \in V \setminus \{t\},\, \propose[v] = \emptyset,\, \refuse[v] \neq * \},
\end{align*}
where $\operatorname{rev}(uv) := vu$ denotes the reversal of a given edge.
Hence, for every vertex $v \in V \setminus \{t\}$, the graph $H_{\propose, \refuse}$ either contains the edge $\propose[v]$, if the proposal pointer is still active, or it contains the reversal $\operatorname{rev}(\refuse[v])$ of the edge $\refuse[v]$, if the refusal pointer is active, or neither of these, if both pointers are inactive. Each edge $e \in E_{H_{\propose, \refuse}}$ has a \emph{residual capacity} $c_f(e)$ depending on the current flow $f$, defined by
\begin{align*}
c_f(e) := \begin{cases}
c(e) - f(e) & \text{if } e \in E,\\
f(e) & \text{if } e = \operatorname{rev}(e') \text{ for some } e' \in E.
\end{cases}
\end{align*}
At the beginning of each iteration of the algorithm, we ensure that no proposal or refusal pointer points to an edge with residual capacity $0$ and that no proposal pointer points to an edge that has already been refused by its head (lines~\ref{alg-line:pointer-update-start}-\ref{alg-line:pointer-update-end}).

\paragraph{Augmenting the flow.}
The algorithm iteratively augments the flow $f$ along an $s$-$t$-path or cycle $W$ in $H_{\propose, \refuse}$ by the bottleneck capacity $\min_{e \in W} c_f(e)$ (lines~\ref{alg-line:augmentation-start}-\ref{alg-line:augmentation-end}). Augmenting a flow $f$ along a path or cycle $W$ by $\Delta$ means that for every $e \in W$, we increase $f(e)$ by $\Delta$ if $e \in E$ and decrease $f(e')$ by $\Delta$ if $e = \operatorname{rev}(e')$ for some $e' \in E$. Note that after the augmentation, $c_f(e) = 0$ for at least one edge $e \in W$, implying that at least one pointer is advanced before the next augmentation. 
Lemma~\ref{lem:path-exists} below shows that an augmenting path or cycle in $H_{\propose, \refuse}$ exists as long as $\propose[s]$ is still active.
The algorithm stops when $\propose[s] = \emptyset$ (lines~\ref{alg-line:termination-start}-\ref{alg-line:termination-end}). 
\medskip

\subsection{Example run of the algorithm}

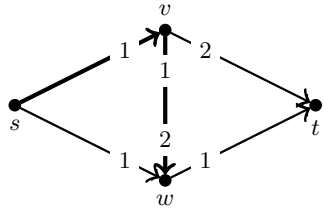
\begin{figure}[t]
	\centering
		\begin{tikzpicture}[scale=1, transform shape,->]
		\pgfmathsetmacro{\b}{2}
		\pgfmathsetmacro{\d}{2}
		\pgfmathsetmacro{\a}{90}
		\pgfmathsetmacro{\c}{45}
		
		\node[vertex, fill = black, label=below:$s$] (s) at (0, 0) {};
		\node[vertex, label=above:$v$] (v) at (2, 1) {};
		\node[vertex, label=below:$w$] (w) at (2, -1) {};
		\node[vertex, fill = black, label=below:$t$] (t) at (4, 0) {};
				
		\draw [pointedline, ultra thick] (s) -- node[edgelabel, near end] {1} (v);
		\draw [pointedline] (s) -- node[edgelabel, near end] {1} (w);
		\draw [pointedline, ultra thick] (v) -- node[edgelabel, near start] {1} node[edgelabel, near end] {2} (w);
		\draw [pointedline] (v) -- node[edgelabel, near start] {2} (t);
		\draw [pointedline] (w) -- node[edgelabel, near start] {1} (t);
		\end{tikzpicture}
\caption{Example instance for illustrating a run of Algorithm~\ref{alg:poly_sf}. Numbers next to the vertices indicate preferences of incident edges. Edge capacities are $c(sv) = c(vw) = 2$ and $c(sw) = c(vt) = c(wt) = 1$. For the algorithm, we choose the arbitrary preference order of the source $s$ to prefer edge $sv$ over $sw$.}
\label{fi:augmenting-algorithm-example}
\end{figure}

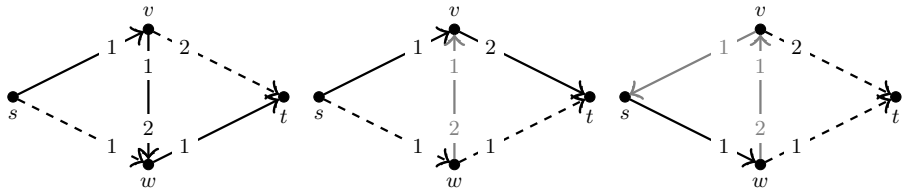
\begin{figure}[t]
\begin{minipage}{.3\textwidth}
	\centering
		\begin{tikzpicture}[scale=0.9, transform shape,->]
		\pgfmathsetmacro{\b}{2}
		\pgfmathsetmacro{\d}{2}
		\pgfmathsetmacro{\a}{90}
		\pgfmathsetmacro{\c}{45}
		
		\node[vertex, fill = black, label=below:$s$] (s) at (0, 0) {};
		\node[vertex, label=above:$v$] (v) at (2, 1) {};
		\node[vertex, label=below:$w$] (w) at (2, -1) {};
		\node[vertex, fill = black, label=below:$t$] (t) at (4, 0) {};
				
		\draw [pointedline] (s) -- node[edgelabel, near end] {1} (v);
		\draw [pointedline, dashed] (s) -- node[edgelabel, near end] {1} (w);
		\draw [pointedline] (v) -- node[edgelabel, near start] {1} node[edgelabel, near end] {2} (w);
		\draw [pointedline, dashed] (v) -- node[edgelabel, near start] {2} (t);
		\draw [pointedline] (w) -- node[edgelabel, near start] {1} (t);
		\end{tikzpicture}
     \end{minipage}\hspace{3mm}
     \begin{minipage}{.3\textwidth}
	\centering
		\begin{tikzpicture}[scale=0.9, transform shape,->]
		\pgfmathsetmacro{\b}{2}
		\pgfmathsetmacro{\d}{2}
		\pgfmathsetmacro{\a}{90}
		\pgfmathsetmacro{\c}{45}
		
		\node[vertex, fill = black, label=below:$s$] (s) at (0, 0) {};
		\node[vertex, label=above:$v$] (v) at (2, 1) {};
		\node[vertex, label=below:$w$] (w) at (2, -1) {};
		\node[vertex, fill = black, label=below:$t$] (t) at (4, 0) {};
				
		\draw [pointedline] (s) -- node[edgelabel, near end] {1} (v);
		\draw [pointedline, dashed] (s) -- node[edgelabel, near end] {1} (w);
		\draw [pointedline, gray] (w) -- node[edgelabel, near start] {2} node[edgelabel, near end] {1} (v);
		\draw [pointedline] (v) -- node[edgelabel, near start] {2} (t);
		\draw [pointedline, dashed] (w) -- node[edgelabel, near start] {1} (t);
		\end{tikzpicture}
     \end{minipage}\hspace{3mm}
          \begin{minipage}{.3\textwidth}
	\centering
		\begin{tikzpicture}[scale=0.9, transform shape,->]
		\pgfmathsetmacro{\b}{2}
		\pgfmathsetmacro{\d}{2}
		\pgfmathsetmacro{\a}{90}
		\pgfmathsetmacro{\c}{45}
		
		\node[vertex, fill = black, label=below:$s$] (s) at (0, 0) {};
		\node[vertex, label=above:$v$] (v) at (2, 1) {};
		\node[vertex, label=below:$w$] (w) at (2, -1) {};
		\node[vertex, fill = black, label=below:$t$] (t) at (4, 0) {};
				
		\draw [pointedline, gray] (v) -- node[edgelabel, near start] {1} (s);
		\draw [pointedline] (s) -- node[edgelabel, near end] {1} (w);
		\draw [pointedline, gray] (w) -- node[edgelabel, near start] {2} node[edgelabel, near end] {1} (v);
		\draw [pointedline, dashed] (v) -- node[edgelabel, near start] {2} (t);
		\draw [pointedline, dashed] (w) -- node[edgelabel, near start] {1} (t);
		\end{tikzpicture}
     \end{minipage}
\caption{The proposal and refusal pointers at the beginning of augmentations 1, 2, and 3, respectively. Proposal pointers are marked by solid black edges, while refusal pointers are the solid gray edges. The dashed edges do not belong to the current set of pointers.}
\label{fi:pointers}
\end{figure}

Before we analyze the algorithm, we illustrate it by running it on the example instance given in Fig.~\ref{fi:augmenting-algorithm-example}. To each augmentation, the set of pointers is drawn in Fig~\ref{fi:pointers}.

\paragraph{Augmentation 1:} Initially, the proposal pointers are set to $\propose[s] = sv$, $\propose[v] = vw$, $\propose[w] = [wt]$, while all refusal pointers are inactive (pointing to $\emptyset$). The graph $H_{\propose, \refuse}$ consists of the edges $sv$, $vw$, and $wt$, which comprise a unique $s$-$t$-path $W_1$. The algorithm augments $f$ along $W_1$ by its bottleneck capacity $1$, yielding the flow $f(sv) = f(vw) = f(wt) = 1$ and $f(sw) = f(vt) = 0$. 

\paragraph{Pointer update:}
Because the residual capacity of $wt$ is $0$, $\textsc{AdvancePointers}(w)$ is called. The procedure advances $\propose[w]$ to the inactive state $\emptyset$ and hence immediately activates $\refuse[w]$ with $\refuse[w] = vw$. 
Because also $\propose[v] = vw$, this pointer is also advanced according to the second criterion of the while loop. It reaches $\propose[v] = vt$.

\paragraph{Augmentation 2:}
With $\propose[s] = sv$, $\refuse[w] = vw$, and $\propose[v] = vt$, the graph $H_{\propose, \refuse}$ consists of the edges $sv$, $\operatorname{rev}(vw) = wv$, and $vt$. The unique $s$-$t$-path $W_2 = \langle s, v, t \rangle$ is chosen, the bottleneck capacity is $c_f(sv) = c_f(vt) = 1$. After augmenting $f$ along $W_2$ by $1$ unit, the new flow is $f(sv) = 2$, $f(vw) = f(vt) = f(wt) = 1$ and $f(sw) = 0$. 

\paragraph{Pointer update:}
Because $c_f(sv) = 0$, the pointer $\propose[s]$ is advanced to $sw$. Because $c_f(vt) = 0$, also $\propose[v]$ is advanced to $\emptyset$ and $\refuse[v]$ gets activated with $\refuse[v] = sv$.

\paragraph{Augmentation 3:} 
With $\propose[s] = sw$, $\refuse[w] = vw$, and $\refuse[v] = sv$, the graph $H_{\propose, \refuse}$ consists of the edges $sw$, $wv$, and $vs$. These edges comprise the cycle $W_3$. The residual capacities are $c_f(sw) = c_f(wv) = 1$ and $c_f(vs) = 2$. Augmenting $f$ along $W_3$ by $1$ unit yields the flow $f(sv) = f(sw) = f(vt) = f(wt) = 1$ and $f(vw) = 0$.

\paragraph{Pointer update:}
Because $c_f(wv) = 0$, the pointer $\refuse[w]$ is updated to $sw$, also triggering an update of $\propose[s]$ that was pointing at the same edge. After advancing $\propose[s]$ it reaches $\emptyset$ and hence the algorithm terminates.

\subsection{Analysis}

In the proof of correctness we utilize the following notation. We say the proposal pointer $\propose[v]$ has \emph{reached} edge $vw$ if $\rank_{v}(\propose[v]) \geq \rank_v(vw)$. We say $\propose[v]$ has \emph{passed} the edge $vw$ if $\rank_{v}(\propose[v]) > \rank_v(vw)$. We use analogous terms for the refusal pointer $\refuse[v]$ with reversed inequality signs, respectively.

We now make a few observations on the behavior of the pointers. We first observe that $\propose[v]$ moves from most-preferred to least-preferred edge and $\refuse[v]$ moves from least-preferred to most-preferred edge, the ranks of the two pointers are non-decreasing or non-increasing, respectively, during the course of the algorithm (note that the lowest rank in $P$ is always higher than the current rank of $\propose[v]$ in line~\ref{alg-line:advance-propose-set} and the highest rank in $R$ is always lower than the current rank of $\refuse[v]$ in line~\ref{alg-line:advance-refuse-set}).

\begin{obs}\label{obs:monotonicity}
Throughout the algorithm, $\rank_{v}(\propose[v])$ never decreases and $\rank_{v}(\refuse[v])$ never increases for any $v \in V \setminus \{t\}$.
\end{obs}

Also, for each vertex, at most one of its two pointers is active at any time, as the refusal pointer is only advanced once the proposal pointer reaches the inactive state.

\begin{obs}\label{obs:active-pointers}
 Throughout the algorithm, for each $v \in V \setminus \{t\}$ either $\refuse[v] = \emptyset$ or $\propose[v] = \emptyset$.
\end{obs}

Finally, we observe that proposal/refusal pointers do not skip any outgoing/incoming edge, respectively. This is due to the construction of $P$ in line~\ref{alg-line:advance-propose-set} and $R$ in line~\ref{alg-line:advance-refuse-set}, which contain every edge that has a rank strictly higher/lower, respectively, than the edge currently pointed at by the pointer.

\begin{obs}\label{obs:advancing-pointers-list}
  Let $uv \in E$.
  \begin{itemize}
  \item If $\rank_u(\propose[u]) < \rank_u(uv)$ before a call of \textsc{AdvancePointers}($u$), then $\rank_u(\propose[u]) \leq \rank_u(uv)$ after that call. 
  \item If $\rank_v(\refuse[v]) > \rank_v(uv)$ before a call of \textsc{AdvancePointers}($v$), then $\rank_v(\refuse[v]) \geq \rank_v(uv)$ after that call.
  \end{itemize}
\end{obs}

We next establish a set of invariants that are useful for analyzing the algorithm.

\begin{lemma}\label{lem:refuse_pointer}
The following invariants hold true for each $uv \in E$ any time the algorithm is in lines~\ref{alg-line:termination-start}-\ref{alg-line:repeat}: 
\begin{enumerate}
\item If $\rank_{v}(\refuse[v]) \leq \rank_{v}(uv)$ then $\propose[u] \neq uv$.\label{inv:proposal-pointer}
\item If $\rank_{v}(\refuse[v]) < \rank_{v}(uv)$ then $f(uv) = 0$.\label{inv:complete-refusal}
\item If $\rank_{u}(\propose[u]) < \rank_{u}(uv)$ then $f(uv) = 0$.\label{inv:flow-increase}
\item 	If $\rank_{u}(\propose[u]) > \rank_{v}(uv)$ then $f(uv) = c(uv)$ or $\rank_{v}(\refuse[v]) \leq \rank_v(uv)$.\label{inv:refused-flow}
\end{enumerate}
\end{lemma}

Note that due to the monotonicity of the pointers, once the premise of invariant \ref{inv:proposal-pointer}, \ref{inv:complete-refusal}, or \ref{inv:refused-flow} is fulfilled for an edge, it will stay this way for the rest of the algorithm.
Intuitively, the invariants state that 
(\ref{inv:proposal-pointer}) a proposal pointer does not point to a refused edge, (\ref{inv:complete-refusal}) once a refusal pointer has passed an edge, the edge carries no flow, (\ref{inv:flow-increase}) an edge can only carry flow after it is reached by its proposal pointer, and (\ref{inv:refused-flow}) after a proposal pointer has passed an edge, the edge is fully saturated until the refusal pointer of its end reaches it.


\proof[of Lemma~\ref{lem:refuse_pointer}]
Invariant~\ref{inv:proposal-pointer}: 
Note that the pointers are only changed in the while loop in lines ~\ref{alg-line:pointer-update-start}-\ref{alg-line:pointer-update-end}.
If $\propose[u] = uv$, then $uv \in E_{H_{\propose, \refuse}}$. Therefore the while loop does not terminate while $\propose[u] = uv$ and $\rank_v(uv) \geq \rank_v(\refuse[v])$.

Invariant~\ref{inv:complete-refusal}: 
Observe the invariant is true after intialization since $f(uv) = 0$.
Note that $f(uv)$ can only increase in line~\ref{alg-line:augmentation-end} when $\propose[u] = uv$. 
In that case, Invariant~\ref{inv:proposal-pointer} ensures that $\rank_v(\refuse[v]) > \rank_v(uv)$.
So the invariant can only become invalid by advancing the pointer $\refuse[v]$ past $uv$.
Consider the first time this happens in the algorithm.
By Observation~\ref{obs:advancing-pointers-list}, this can only happen 
with a call of \textsc{AdvancePointers}($v$) when $\refuse[v] = uv$.
But then $\propose[v] = \emptyset$ by Observation~\ref{obs:active-pointers} and therefore the call of \textsc{AdvancePointers}($v$) can only be triggered by the condition $c_f(vu) = 0$ of the while loop. But this implies $f(uv) = 0$, so the invariant did not become invalid.
 
Invariant~\ref{inv:flow-increase}: Initially, $f(uv) = 0$. The flow can only increase when $uv$ is part of an augmenting path or cycle in line~\ref{alg-line:augmentation-end}. This can only happen while $\propose[u] = uv$ by construction of $E_{H_{\propose,\refuse}}$. Because $\rank_u(\propose[u])$ is non-decreasing, $\rank_u(\propose[u]) \geq \rank_u(uv)$ is true at any time after the first increase of $f(uv)$.
 \smallskip
 
Invariant~\ref{inv:refused-flow}: This invariant is true initially because $\refuse[v] = \emptyset$. It can only lose its validity by advancing $\propose[u]$ or decreasing $f(uv)$. By Observation~\ref{obs:advancing-pointers-list}, $\propose[u]$ can only pass $uv$ when $\textsc{AdvancePointers}(u)$ is called in line~\ref{alg-line:pointer-update-end} while $\propose[u] = uv$. This call can be triggered because $\rank_v(\refuse[v]) \leq \rank_v(uv)$ or because $c_f(uv) = 0$ (implying $f(uv) = c(uv)$). In either case, the invariant is not violated. The flow on $f(uv)$ can only decrease when $\operatorname{rev}(uv) \in W \subseteq E_{H_{\propose,\refuse}}$. By definition, this can only happen if $\refuse[v] = uv$, which is already enough to fulfill the invariant.\qed

\medskip

With the following lemma, we show that, at the beginning of each iteration, the algorithm can actually find an $s$-$t$-path or cycle.

\begin{lemma}\label{lem:path-exists}
Each time the algorithm reaches line~\ref{alg-line:augmentation-start}, the graph $H_{\propose, \refuse}$ contains an $s$-$t$-path or a cycle.
\end{lemma}

\proof
Consider any $v \in V \setminus \{s, t\}$ at any time the algorithm reaches line~\ref{alg-line:augmentation-start}. 
We show that if $v$ has an incoming edge in $H_{\propose, \refuse}$, then it also has an outgoing edge in $H_{\propose, \refuse}$. Note that by definition of $E_{H_{\propose, \refuse}}$, the only situation in which $v$ has no outgoing edge is when $\refuse[v] = *$.

Let $uv \in E_{H_{\propose, \refuse}}$ be an incoming edge of $v$.
This implies that either $uv \in E$ and $\propose[u] = uv$ or $vu \in E$ and $\refuse[u] = vu$ by definition of $H_{\propose, \refuse}$. 

If $\propose[u] = uv$, Invariant~\ref{inv:proposal-pointer} of Lemma~\ref{lem:refuse_pointer} ensures that $\rank_v(\refuse[v]) > \rank_v(uv)$ and hence $\refuse[v] \neq *$. Therefore $v$ has an outgoing edge in $H_{\propose, \refuse}$.

If $vu \in E$ and $\refuse[u] = vu$, the termination criterion of the while loop (lines~\ref{alg-line:pointer-update-start}-\ref{alg-line:pointer-update-end}) guarantees  
$f(vu) = c_f(\operatorname{rev}(uv)) > 0$.
Hence, by flow conservation, $v$ must also have an incoming edge $u'v \in E$ with $f(u'v) > 0$. By Invariant~\ref{inv:complete-refusal} of Lemma~\ref{lem:refuse_pointer}, this implies $\refuse[v] \neq *$.

Thus every non-terminal vertex with an incoming edge also has an outgoing edge.
Now observe that $\propose[s] \neq \emptyset$ ensures that $s$ also has an outgoing edge in $H_{\propose, \refuse}$. Thus, we can start a walk at $s$ and extend it until we visit a vertex as second time, closing a cycle, or until we reach $t$ having found an $s$-$t$-path. This concludes the proof of the lemma.
\qed

\begin{theorem}
	Algorithm~\ref{alg:poly_sf} computes a stable flow in polynomial time.
\end{theorem}

\proof 
%

We first show that the algorithm indeed computes a stable flow. Assume by contradiction there is a walk $W = \langle v_1, v_2, \dots, v_k \rangle$ blocking~$f$. We use the previously established invariants to prove the following claim.

\begin{claim}
For every $i \in \{1, \dots, k-1\}$, the pointer $\propose[v_i]$ has passed $v_iv_{i+1}$, i.e., $\rank_{v_i}(\propose[v_i]) > \rank_{v_i}(v_iv_{i+1})$.
\end{claim}
\myproof
We show the claim by induction on $i$.
First consider the case $i = 1$.
Due to point~2 in Definition~\ref{def:sf}, either $v_1 = s$ or $\rank_{v_1} (v_1v_2) < \rank_{v_1} (v_1w)$ for some $v_1w \in E$ with $f(v_1w) > 0$. In the former case, $\propose[s]$ has passed $v_1v_2$ as the termination criterion of the algorithm implies $\propose[s] = \emptyset$. In the latter case, $f(v_1w) > 0$ implies that $\propose[v_1]$ has at least reached $v_1w$ by Invariant~\ref{inv:flow-increase} of Lemma~\ref{lem:refuse_pointer} and thus it has passed $v_1v_2$.

Now consider any $i \in \{2, \dots, k-1\}$. Note that by induction hypothesis $\propose[v_{i-1}]$ has passed $v_{i-1}v_{i}$. Furthermore $f(v_{i-1}v_{i}) < c(v_{i-1}v_{i})$ because no edge of $W$ is saturated. Hence, Invariant~\ref{inv:refused-flow} of Lemma~\ref{lem:refuse_pointer} implies that $\refuse[v_i]$ must have reached $v_{i-1}v_{i}$. In particular, $\refuse[v_i] \neq \emptyset$ and hence $\propose[v_i] = \emptyset$ by Observation~\ref{obs:active-pointers}, implying $\propose[v_i]$ has passed all edges. This completes the induction and proves the claim. \myqed\smallskip

Now consider $v_k$, the last vertex of $W$.
Note that, due to the claim above, $\propose[v_{k-1}]$ has passed $v_{k-1}v_k$.
Furthermore, $f(v_{k-1}v_{k}) < c(v_{k-1}v_{k})$ as the blocking walk $W$ is unsaturated. Hence, by Invariant~\ref{inv:refused-flow} of Lemma~\ref{lem:refuse_pointer}, $\refuse[v_k]$ has reached $v_{k-1}v_{k}$, i.e., $\rank_{v_k}(\refuse[v_k]) \leq \rank_{v_k}(v_{k-1}v_k)$.

Observe that this implies $\rank_{v_k}(\refuse[v_k]) < \infty = \rank_t(\refuse[t])$ and therefore $v_k \neq t$ (remember that $\refuse[t] = \emptyset$ never changes).
Now consider any $uv_k \in E$ with $\rank_{v_k} (v_{k-1}v_k) < \rank_{v_k} (uv_k)$. 
Then $\rank_{v_k}(\refuse[v_k]) \leq \rank_{v_k}(v_{k-1}v_{k}) < \rank_{v_k} (uv_k)$ implies 
$f(uv_k) = 0$ by Invariant~\ref{inv:complete-refusal} of Lemma~\ref{lem:refuse_pointer}.
Therefore $W$ does not dominate $f$ at the end, i.e., it does not fulfill point~3 of Definition~\ref{def:sf}. Thus $W$ is not a blocking walk and the returned flow $f$ is stable.

We now turn to the running time. 
Note that in every iteration of the while loop (lines~\ref{alg-line:pointer-update-start}-\ref{alg-line:pointer-update-end}), a pointer of a vertex is advanced. Thus the total number of iterations of the while loop throughout the whole algorithm is bounded by $2|E|$ by monotonicity of the pointers and the fact that each edge appears in at most two preference lists. 
Since every vertex has at most one incoming and one outgoing edge in $H_{\propose, \refuse}$ by construction, finding edges violating the termination criterion of the loop can be done in time $\mathcal{O}(|V|)$.
The same is true for finding an augmenting path or cycle in line~\ref{alg-line:augmentation-start}.
As after each augmentation, the residual capacity of at least one edge drops to $0$, at least one pointer is advanced in line~\ref{alg-line:pointer-update-end} between any two augmentations, limiting the number of augmentations by $2|E|$.
Hence the total running time of the algorithm is bounded by $\mathcal{O}(|E||V|)$.
 We remark that a more sophisticated implementation using the dynamic-tree data structure can reduce this running time to $\mathcal{O}(|E|\log|V|)$. However, since our primary aim in this article is to provide new and simple approaches, we omit further investigation of this complication. \qed

\section{Stable flows with restricted intervals}
\label{sec:re_sf}

Various stable matching problems have been tackled under the assumption that restricted edges are present in the graph~\cite{DFFS03,FIM07}. A restricted edge can be \emph{forced} or \emph{forbidden}, and the aim is to find a stable matching that contains all forced edges, while it avoids all forbidden edges. 
Such edges correspond to transactions that are particularly desirable or undesirable from a social welfare perspective, but it is undesirable or impossible to push the participating agents directly to use or avoid the edges. We thus look for a stable solution in which the edge restrictions are met voluntarily.

A natural way to generalize the notion of a restricted edge to the stable flow setting is to require the flow value on any given edge to be within a certain interval.
To this end, we introduce a \emph{lower} and an \emph{upper bound} function. 

\begin{pr} \textsc{sf restricted}\ \\
	\inp $\mathcal{I} = (D, c, r, \mathfrak{l}, \mathfrak{u})$; an \textsc{sf} instance $(D, c, r)$, a lower bound function $\mathfrak{l}: E \rightarrow \mathbb{R}_{\geq 0}$ and an upper bound function $\mathfrak{u}: E \rightarrow \mathbb{R}_{\geq 0}$. \\
	\ques Is there a stable flow $f$ so that $\mathfrak{l}(uv) \leq f(uv) \leq \mathfrak{u}(uv)$ for all $uv \in E$?
\end{pr}

Note that in the above definition, the upper bound $\mathfrak{u}$ does not affect blocking walks, i.e., a blocking walk can use edge $uv$, even if $f(uv) = \mathfrak{u}(uv) < c(uv)$ holds. 
In particular, it is not without loss of generality to assume $c(uv) = \mathfrak{u}(uv)$ for all edges $uv$, as decreasing $c(uv)$ may enlarge the set of stable flows.

In the following, we describe a polynomial algorithm that finds a stable flow with restricted intervals or proves its nonexistence. We start with an instance modification step in Section~\ref{sec:sf_forbidden_reduction}. Then we prove that restricted intervals can be handled by small network modifications that reduce the problem to the unrestricted version of \textsc{sf}. We show this separately for the case where only forced edges occur, which we call \textsc{sf forced}, in Section~\ref{sec:forced} and for the case where only forbidden edges occur, called \textsc{sf forbidden}, in Section~\ref{sec:forbidden}. It is straightforward to see that these two results can be combined to solve the general version of \textsc{sf restricted}.

We mention that it is also possible to solve \textsc{sf restricted} by transforming the instance first into a weighted \textsc{sf} instance, and then into a weighted stable allocation instance, both solvable in $\mathcal{O}(|E|^2 \log|V|)$ time~\cite{DM10}. The advantages of our method are that it can be applied directly to the \textsc{sf restricted} instance and it also gives us insights to solving the stable roommate problem with restricted edges directly, as pointed out at the end of Sections~\ref{sec:forced} and~\ref{sec:forbidden}. Moreover, our running time is only~$\mathcal{O}(|P||E| \log|V|)$, where $P$ is the set of edges with $\mathfrak{u}(uv) < c(uv)$.

\subsection{Problem simplification}
\label{sec:sf_forbidden_reduction}
\textsc{sf restricted} generalizes the natural notion of requiring flow to use an edge to its full capacity (by setting $\mathfrak{l}(uv) = c(uv)$) and of requiring flow not to use an edge at all (by setting $\mathfrak{u}(uv) = 0$), which corresponds to the traditional cases of forced and forbidden edges. 
In fact, it turns out that any given instance of \textsc{sf restricted} can be transformed into an equivalent instance in which \mbox{$\mathfrak{l}(uv), \mathfrak{u}(uv) \in \left\{ 0,c(uv) \right\}$} for all~$uv \in E$. 

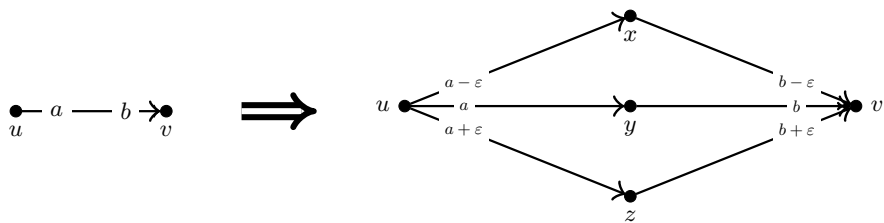
\begin{figure}
	\begin{minipage}{0.4\textwidth}
		\begin{tikzpicture}[scale=1, transform shape,->]
		\pgfmathsetmacro{\b}{2}
		\pgfmathsetmacro{\d}{2}
		
		\node[vertex, label=below:$u$] (u) at (0, 0) {};
		\node[vertex, label=below:$v$] (v) at (\b, 0) {};
		
		\draw [pointedline] (u) -- node[edgelabel, near start] {$a$} node[edgelabel, near end] {$b$} (v);
		\draw [arrow](3,0)-- (4,0);
		\end{tikzpicture}
	\end{minipage}\begin{minipage}{0.6\textwidth}
	\begin{tikzpicture}[scale=1, transform shape]
	\pgfmathsetmacro{\b}{3}
	\pgfmathsetmacro{\d}{1.2}
	
	\node[vertex, label=left:$u$] (u) at (0, 0) {};
	\node[vertex, label=right:$v$] (v) at (2*\b, 0) {};
	\node[vertex, label=below:$x$] (x) at (\b, \d) {};
	\node[vertex, label=below:$y$] (y) at (\b, 0) {};
	\node[vertex, label=below:$z$] (z) at (\b, -\d) {};
	
	\draw [pointedline] (u) -- node[edgelabel, near start, scale=0.7] {$a-\varepsilon$} (x);
	\draw [pointedline] (u) -- node[edgelabel, near start, scale=0.7] {$a$} (y);
	\draw [pointedline] (u) -- node[edgelabel, near start, scale=0.7] {$a+\varepsilon$}(z);
	
	\draw [pointedline] (x) -- node[edgelabel, near end, scale=0.7] {$b-\varepsilon$} (v);
	\draw [pointedline] (y) -- node[edgelabel, near end, scale=0.7] {$b$} (v);
	\draw [pointedline] (z) -- node[edgelabel, near end, scale=0.7] {$b+\varepsilon$} (v);
	\end{tikzpicture}
\end{minipage}
\caption{Splitting an edge with lower and upper bounds. Due to the preferences, capacities and bounds defined on the modified instance, the first $\mathfrak{l}(uv)$ units of flow will saturate $\langle u, x, v \rangle$, then, the coming $\mathfrak{u}(uv) - \mathfrak{l}(uv)$ units of flow will saturate $\langle u, y, v \rangle$, and the remaining $c(uv) - \mathfrak{u}(uv)$ units of flow will use $\langle u, z, v \rangle$.}
\label{fi:split}
\end{figure}

First observe that if $\mathfrak{l}(uv) > \mathfrak{u}(uv)$ for some $uv \in E$, then \textsc{sf restricted} trivially has no solution. 
Therefore, we henceforth assume $\mathfrak{l}(uv) \leq \mathfrak{u}(uv)$ for all $uv \in E$. We further execute the following technical change to the instance in order to obtain an equivalent instance with the desired properties. As shown in Fig.~\ref{fi:split}, we substitute each edge $uv \in E$ with three parallel paths (to avoid parallel edges): $\langle u, x, v \rangle, \langle u, y, v \rangle$ and~$\langle u, z, v \rangle$. While $uy$ and $yv$ take over the rank of $uv$, $ux$ and $xv$ are ranked just above, $uz$ and $zv$ are ranked just below $uy$ and~$yv$. The capacities and bounds of the introduced edges are as follows.
\vspace{-0.4cm}

\begin{alignat*}{7}
\mathfrak{l}(ux) & \;=\; \mathfrak{l}(xv) && \;=\; \mathfrak{u}(ux) && \;=\; \mathfrak{u}(xv) && \;=\; c(ux) & \;=\; c(xv) & \;=\; \mathfrak{l}(uv)\\
\mathfrak{l}(uy) & \;=\; \mathfrak{l}(yv) && \;=\; 0 \\
\mathfrak{u}(uy) & \;=\; \mathfrak{u}(yv) && \;=\; c(uy) && \;=\; c(yv) && \;=\; \mathfrak{u}(uv) - \mathfrak{l}(uv)\hspace*{-1.5cm}\\
\mathfrak{l}(uz) & \;=\; \mathfrak{l}(zv) && \;=\; \mathfrak{u}(uz) && \;=\; \mathfrak{u}(zv) && \;=\; 0\\
c(uz)& \;=\; c(zv) && \;=\; c(uv) - \mathfrak{u}(uv)\hspace*{-1.5cm}
\end{alignat*}

In words, we split each edge $uv$ with lower and upper bounds into three paths: the first path $\langle u, x, v \rangle$ requires an amount of flow exactly equal to its capacity~$\mathfrak{l}(uv)$, 
the middle path $\langle u, y, v \rangle$ has capacity $\mathfrak{u}(uv) - \mathfrak{l}(uv)$ and is unrestricted,
the last path $\langle u, z, v \rangle$ with capacity $c(uv) - \mathfrak{u}(uv)$ must not carry any flow. 

Note that we can map any flow $f$ in original graph to a flow $f'$ in the modified graph by splitting the flow on each edge $uv$ into three parts, setting $f'(ux) = f'(xv) = \min \{f(uv),\mathfrak{l}(uv) \}$, $f'(uy) = f'(yv) = \min \{\max \{f(uv) - \mathfrak{l}(uv), 0\}, \mathfrak{u}(uv) \}$, and $f'(uz) = f'(zv) = \max \{f(uv) - \mathfrak{u}(uv), 0 \}$. 
Conversely, every flow $f'$ in the modified instance induces a flow $f$ in the original instance, simply by aggregating the flow values on the three paths, i.e., setting $f(uv) = f(ux) + f(uy) + f(uz)$.

Note that different flows in the modified instance can map to the same flow $f$ in the original network, but it is easy to check that if $f$ is stable, only a unique stable flow in the modified instance maps to $f$. Thus there is a one-to-one correspondence between stable flows in the original instance and in the modified instance. 
Furthermore, it is straightforward to check that $f$ respects the bounds $\mathfrak{l}$ and $\mathfrak{u}$ in the original instance if and only if $f'$ does the same in the modified instance. The modified instance is thus equivalent to the original instance.

\begin{remark}
\label{rem:sf_forbidden_reduction}
Note that the encoding size of the modified instance is within a constant factor of the instance size of the original instance.
More precisely, the number of edges in the new instance is $6|E|$ and the number of nodes in the new instance is $|V| + 3|E|$, where $V$ and $E$ are the sets of vertices and edges of the original instance, respectively.
Also the set $P$ of edges with $\mathfrak{u}(e) < c(e)$ only grows by a factor of $2$.
Note that because we assumed the original graph to be simple and connected, $|V| - 1 \leq |E| \leq |V|^2$ and therefore $\log(|V| + 3|E|) = \mathcal{O}(\log |V|)$. Therefore the asymptotic running time of $\mathcal{O}(|P| |E| \log |V|)$ which we will establish for our algorithm on the modified instance is the same for the original instance.
\end{remark}

Henceforth, we will assume that our instances are of this form and use the notation $Q := \left\{ uv \in E \st \mathfrak{l}(uv) = c(uv) \right\}$ and $P := \left\{ uv \in E \st \mathfrak{u}(uv) = 0 \right\}$ for the sets of forced and forbidden edges, respectively.

\subsection{Forced edges}
\label{sec:forced}

In this section we consider an instance of \textsc{sf restricted} where $P = \emptyset$. As mentioned earlier, we call this problem \textsc{sf forced}. In Section~\ref{se:singleforced} we show how to deal with the case $|Q| = 1$ by reducing the corresponding \textsc{sf forced} instance with a single forced edge to an instance of \textsc{sf} without forced edges. Then, in Section~\ref{se:multipleforced}, we argue that the same technique can be applied to multiple forced edges simultaneously. At last, in Section~\ref{se:forcedmatching} we elaborate on the application of our technique for stable matching instances.

\subsubsection{A single forced edge}
\label{se:singleforced}

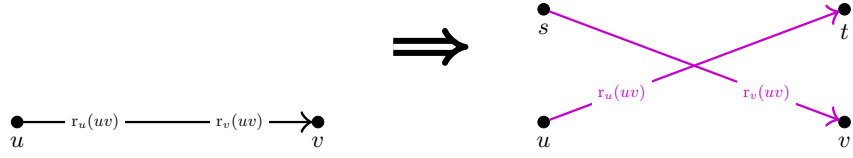
\begin{figure}[t]
\centering
\begin{tikzpicture}[scale=1, transform shape]

\node[vertex, label=below:$u$] (a) at (0, 0) {};
\node[vertex, label=below:$v$] (b) at (4, 0) {};
\node[vertex, fill = black, label=below:$s$] (s) at (7, 1.5) {};
\node[vertex, label=below:$u$] (d) at (7, 0) {};
\node[vertex, label=below:$v$] (e) at (11,0) {};
\node[vertex, fill = black, label=below:$t$] (t) at (11,1.5) {};

\draw [pointedline] (a) -- node[edgelabel, near start, scale=0.7] {$\rank_u(uv)$} node[edgelabel, near end, scale=0.7] {$\rank_v(uv)$} (b);
\draw [pointedline, MyPurple] (s) -- node[edgelabel, near end, scale=0.7] {$\rank_v(uv)$} (e);
\draw [pointedline, MyPurple] (d) -- node[edgelabel, near start, scale=0.7] {$\rank_u(uv)$} (t);

\draw [arrow](5,1)-- (6,1);
\end{tikzpicture}
\caption{Substituting forced edge $uv$ by edges $sv$ and $ut$ in $D'$.}
\label{fi:forcedsf}
\end{figure}

Let us first consider a single forced edge $uv$. We modify graph $D$ to derive a graph~$D'$. The modification consists of deleting the forced edge $uv$ and introducing two new edges $sv$ and $ut$ to substitute it. Both new edges have capacity $c(uv)$ and take over $uv$'s rank on $u$'s and on $v$'s preference lists, respectively, as shown in Fig.~\ref{fi:forcedsf}. The rest of $D$ remains unchanged in~$D'$.

In Lemma~\ref{le:singleforced} we show that flows saturating $uv$ in $D$ are equivalent to flows saturating both $sv$ and $ut$ in~$D'$. Then we refer to the extension of the Rural Hospitals Theorem (Theorem~\ref{same_flow_value}) to solve the latter problem.
	
\begin{lemma}
\label{le:singleforced}
    Let $f$ be a flow in $D$ with $f(uv) = c(uv)$. Let $f'$ be the flow in $D'$ derived by setting $f'(sv) = f'(ut) = f(uv)$ and $f'(e) = f(e)$ for all $e \in E \setminus \{uv\}$. Then $f$ is stable if and only if $f'$ is stable.
\end{lemma}

\proof
We prove this lemma by showing that walks blocking $f$ also block $f'$ and vice versa. We first observe that the set of edges not saturated by $f$ in $D$ is the same as the set of edges not saturated by $f'$ in $D'$. This is because $uv$ is saturated by $f$, and therefore $ut, sv$ are saturated by $f'$, and all other edges are present in both graphs with identical capacities and flow values, respectively.
Note that this implies the set of walks in $D$ not saturated by $f$ and the set of walks in $D'$ not saturated by $f'$ is the same.

Now consider any node $u' \in V$ and any number $r > 0$. Observe that there is an edge $u'v'$ in $D$ with $\rank_{u'}(u'v') = r$ and $f(u'v') > 0$ if and only if there is $u'v''$ in $D'$ with $\rank_{u'}(u'v'') = r$ and $f'(u'v'') > 0$ (either $u'v'$ itself is in $D'$ or $u'v' = uv$, in which case $u'v'' = ut$ fulfills the requirement). Therefore an unsaturated walk $W$ in $D$ dominates $f$ at the start if and only if it dominates $f'$ at the start.
A symmetric argument holds for dominance at the end of an unsaturated walk. 
This implies that any blocking walk for $f$ in $D$ is a blocking walk for $f'$ in $D'$ and vice versa. \qed
\medskip
%

Checking the existence of a flow in $D'$ that saturates both $sv$ and $ut$ can be done by finding any stable flow in~$D'$. This is because Theorem~\ref{same_flow_value} guarantees that all stable flows have the same value on any edge incident to $s$ or~$t$.

\subsubsection{Multiple forced edges}
\label{se:multipleforced}

We observe that we can replace all edges in $Q$ one after the other, applying Lemma~\ref{le:singleforced} inductively on the resulting graph. This yields the following theorem.

\begin{theorem}\label{thm:multiple-forced-edges}
  Let $D_Q$ be the graph obtained from $D$ when replacing each edge in $uv \in Q$ by edges $ut$ and $sv$ with same rank and capacity. Let $\bar{Q}$ be the set of newly added edges in $D_Q$.
  Let $f$ be a flow in $D$ saturating all edges in $Q$.
  Then $f$ is stable if and only if the corresponding flow $f'$ in $D_Q$ obtained by setting $f'(sv) = f'(ut) = f(uv)$ for all $uv \in Q$ and $f(e) = f'(e)$ for all $e \in E \setminus Q$ is stable.
\end{theorem}

In fact, the Rural Hospitals Theorem (Theorem~\ref{same_flow_value}) guarantees that either all stable flows in $D_Q$ saturate all edges in $\bar{Q}$ or none does. Thus we can solve \textsc{sf forced} by a single stable flow computation in $D_Q$.

\begin{theorem}
\textsc{sf forced} can be solved in time $\mathcal{O}(|E| \log |V|)$.
\end{theorem}

\proof
As $D_Q$ contains at most twice as many edges as $D$, we can compute a stable flow $f'$ in $D_Q$ in time $\mathcal{O}(|E| \log |V|)$, as discussed at the end of Section~\ref{polalg_sf}. If $f'(sv) = f'(ut) = c(uv)$ for all $uv \in Q$, the corresponding flow in $D$ with $f(uv) = f'(sv)$ is a stable flow in $D$ saturating all edges in $Q$. 
Now assume $f'(sv) < c(uv)$ or $f'(ut) < c(uv)$ for some $uv \in Q$. Then by Theorem~\ref{same_flow_value}, any stable flow in $D_Q$ has this property. Hence, no stable flow in $D$ saturates all edges in~$Q$.\qed

\subsubsection{Stable matchings with forced edges}
\label{se:forcedmatching}

We shortly discuss the case of forced edges in stable matching instances. Notice that our observations are valid in the so-called stable roommates setting, where the underlying graph is not bipartite. The definition of a blocking edge is exactly the same as in the classical bipartite case. An edge $uv \notin M$ blocks $M$ if both $u$ and $v$ prefer each other to their respective partners in~$M$.

\begin{problem} \textsc{sr forced}\ \\
	\inp $\mathcal{I} = (G, r, Q)$; a graph $G$ (not necessarily bipartite), the preference ordering $r$ of vertices, and a set of forced edges~$Q$. \\
	\ques Is there a stable matching covering all edges in~$Q$?
\end{problem}

The technique described above provides a fairly simple method for solving \textsc{sr forced}, because the Rural Hospitals Theorem holds for the stable roommates problem as well~\cite[Theorem 4.5.2]{GI89}. After deleting each forced edge $uw \in Q$ from the graph, we add $uw_s$ and $u_tw$ edges to each of the pairs, where $w_s$ and $u_t$ are newly introduced vertices. These edges take over the rank of~$uw$. Unlike in \textsc{sf}, here we need to introduce two separate dummy vertices to each forced edge, simply due to the matching constraints. There is a stable matching containing all forced edges if and only if an arbitrary stable matching covers all of these new vertices $w_s$ and~$u_t$. The proof for this is analogous to that of Lemma~\ref{le:singleforced}.

The running time of this algorithm is~$\mathcal{O}(|E|)$, since it is sufficient to construct a single stable solution in an instance with at most $2|V|$ vertices. More vertices cannot occur, because in  a matching problem more than one forced edge incident to a vertex immediately implies infeasibility. Notice that solving \textsc{sr forced} has the same time complexity $\mathcal{O}(|E|)$ as solving the stable roommates problem without any restriction on the edges. 

\subsection{Forbidden edges}
\label{sec:forbidden}
 
In order to handle \textsc{sf forbidden}, we present here an argumentation of the same structure as in the previous section. In Section~\ref{se:singleforbidden}, we show how to solve the problem of stable flows with a single forbidden edge by solving two instances on two different extended networks. Then, in Section~\ref{se:multipleforbidden} we show how these constructions can be used to obtain an algorithm for the case of multiple forbidden edges. Finally, in Section~\ref{se:matchingforbidden} we discuss the implication of our results to stable matching instances.

Now we introduce some notation used in this section. We remind the reader that $P$ is the set forbidden edges, where $\mathfrak{l}(e) = c(e)$. For $e = uv \in P$, we define edges $e^+ = sv$ and $e^- = ut$. We set $c(e^+) = \varepsilon > 0$ and set $r_v(e^+) = r_v(e) - \varepsilon$, i.e., $e^+$ occurs on $v$'s preference list exactly before~$e$. Likewise, we set $c(e^-) = \varepsilon$ and $r_u(e^-) = r_u(e) - \varepsilon$,  i.e., $e^-$ occurs on $u$'s preference list exactly before~$e$.
For $F \subseteq P$ we define $E^+(F) := \{e^+ : e \in F\}$ and $E^-(F) := \{e^- : e \in F\}$.

\subsubsection{A single forbidden edge}
\label{se:singleforbidden}
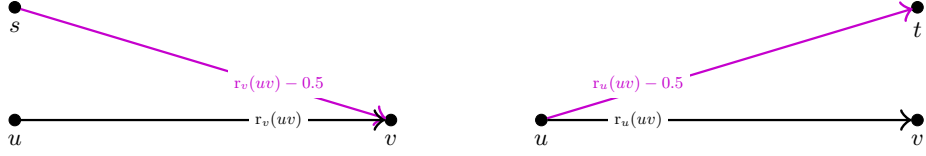
\begin{figure}[t]
\centering
\begin{tikzpicture}[scale=1, transform shape]

\node[vertex, label=below:$u$] (a) at (-1, 0) {};
\node[vertex, label=below:$v$] (b) at (4, 0) {};
\node[vertex, fill = black, label=below:$s$] (s) at (-1, 1.5) {};
\node[vertex, label=below:$u$] (d) at (6, 0) {};
\node[vertex, label=below:$v$] (e) at (11,0) {};
\node[vertex, fill = black, label=below:$t$] (t) at (11,1.5) {};

\draw [pointedline, MyPurple] (s) -- node[edgelabel, near end, xshift=-2mm, yshift=1mm, scale=0.7] {$\rank_v(uv) - 0.5$} (b);
\draw [pointedline] (a) -- node[edgelabel, near end, xshift=-2mm, yshift=0mm, scale=0.7] {$\rank_v(uv)$} (b);
\draw [pointedline, MyPurple] (d) -- node[edgelabel, near start, xshift=0mm, yshift=1mm, scale=0.7] {$\rank_u(uv) - 0.5$} (t);
\draw [pointedline] (d) -- node[edgelabel, near start, xshift=0mm, yshift=0mm, scale=0.7] {$\rank_u(uv)$} (e);

\end{tikzpicture}
\caption{Adding edges $e_0^+ = sv$ in $D^+$ and $e_0^- = ut$ in $D^{-}$ to forbidden edge~$E = uv$.}
\label{fi:forbiddensf}
\end{figure}

Assume that $P = \left\{ e_0 \right\}$ for a single edge $e_0$. First we present two modified instances that will come handy when solving \textsc{sf forbidden}. The first is the graph~$D^+$, which we obtain from $D$ by adding the edge $e_0^+$ to~$E$. Similarly, we obtain the graph $D^{-}$ by adding $e_0^-$ to $E$. Both graphs are illustrated in Fig.~\ref{fi:forbiddensf}.

In the following, we characterize \textsc{sf forbidden} instances with the help of $D^{+}$ and~$D^{-}$. Our claim is that \textsc{sf forbidden} in $D$ has a solution if and only if there is a stable flow $f^+$ in $D^+$ with~$f^+(e^+)=0$ or there is a stable flow $f^-$ in $D^-$ with~$f^-(e^-)=0$. These existence problems can be solved easily in polynomial time, since all stable flows have the same value on edges incident to terminal vertices by Theorem~\ref{same_flow_value}.

We start with a straightforward observation, which follows from the fact that the deletion of an edge that does not carry any flow in a stable flow neither affects flow conservation nor can create blocking walks.

\begin{obs}
\label{ob:delete}
	If $f(e) = 0$ for an edge $e \in E$ and stable flow $f$ in $D$, then $f$ remains stable in $D - e$ as well.
\end{obs}

Now we are ready to prove the correctness of our transformation. 

\begin{lemma}
\label{le:single_forbidden}
  Let $f$ be a flow in $D = (V, E)$ with $f(e_0) = 0$. Then $f$ is a stable flow in $D$ if and only if at least one of the following properties hold:
	\begin{enumerate}[Property~1:]
		\item The flow $f^+$ with $f^+(e) = f(e)$ for all $e \in E$ and $f^+(e_0^+)=0$ is stable in $(V, D^{+})$.
		\item The flow $f^-$ with $f^-(e) = f(e)$ for all $e \in E$ and $f^-(e_0^-)=0$ is stable in $(V, D^{-})$.
	\end{enumerate}
\end{lemma}

\proof
  Sufficiency of any of the two properties follows immediately from Observation~\ref{ob:delete} by deletion of $e_0^+$ or $e_0^-$, respectively, since there edges carry zero flow.

  To see necessity, assume that $f$ is a stable flow in $D$. By contradiction assume that neither $f^+$ nor $f^-$ is stable. Then there is a blocking walk $W^+$ for $f^+$ and a blocking walk $W^-$ for $f^-$. Since $W^+$ is not a blocking walk for $f$ in $D$, it must contain $e_0^+$. This is only possible if $W^+$ starts with $e_0^+$, because $e_0^+$ starts at a terminal vertex. Similarly, since $W^-$ is not a blocking walk for $f$ in $D$, it must end with $e_0^-$. Let $W'^+ := W^+ \setminus \{e_0^+\}$ and $W'^- := W^- \setminus \{e_0^-\}$. Consider the concatenation $W := W'^- \circ e_0 \circ W'^+$. Note that $W$ is an unsaturated walk in~$D$. If $W'^- \neq \emptyset$, then $W$ starts with the same edge as $W^-$ and thus dominates $f$ at the start. If $W'^- = \emptyset$, then $W$ starts with $e_0$, which dominates any flow-carrying edge dominated by $e_0^-$, and hence it dominates $f$ at the start also in this case. By analogous arguments it follows that $W$ also dominates $f$ at the end. Hence $W$ is a blocking walk, contradicting the stability of $f$. We conclude that at least one of Properties~1 or~2 must be true if $f$ is stable. \qed
  
This method can be used to solve \textsc{sf forbidden} if $|P| = 1$, by simply computing stable flows $f^+$ in $D^+$ and $f^-$ in $D^-$. 
Note that by the extension of the Rural Hospitals Theorem (Theorem~\ref{same_flow_value}), the flow values $f^+(e_0^+)$ and $f^-(e_0^-)$ do not depend on the choice of $f^+$ and $f^-$, since they are the same for all stable flows in an instance. If $f^+(e_0^+) = 0$ or $f^-(e_0^-) = 0$, then we have found a stable flow in $f$ avoiding the forbidden edge $e_0$. On the other hand, if the flow value is positive in both cases, there is no stable flow avoiding~$e_0$.

\subsubsection{Multiple forbidden edges}
\label{se:multipleforbidden}

For $|P| > 1$, Lemma~\ref{le:single_forbidden} guarantees that we can add either $e^+$ or $e^-$ for each forbidden edge $e \in P$ without destroying any stable flow avoiding the forbidden edges. However, it is not straightforward to decide for which forbidden edges to add $e^+$ and for which to add $e^-$. 
Simply checking the two properties in Lemma~\ref{le:single_forbidden} and creating either a $D^-$ or $D^+$ graph for each forbidden edge in an arbitrary order does not lead to correct results, since the modification steps can impact each other. It is possible that the forbidden edge checked first allows for both $D^-$ and $D^+$, and it turns out at  a later forbidden edge that only one of these two choices can be combined with network modifications induced when tackling other forbidden edges, as the following example reveals. The same example demonstrates that adding both $e^+$ and $e^-$ to all forbidden edges at the same time might lead to an instance that admits no stable flow avoiding all added edges, even though a stable flow avoiding all forbidden edges exists in the original instance. After the example we describe how to resolve this issue and obtain a polynomial time algorithm for \textsc{sf forbidden}.

\begin{figure}[t]
\centering
\begin{tikzpicture}[scale=1, transform shape]

\node[vertex, label=right:$v_3$] (j_3) at (1, 4) {};
\node[vertex, label=right:$v_2$] (j_2) at (-4, 4) {};
\node[vertex, label=left:$v_1$] (j_1) at (-9, 4) {};

\node[vertex, label=left:$u_1$] (m_1) at (-9, 0) {};
\node[vertex, label=left:$u_2$] (m_2) at (-4, 0) {};
\node[vertex, label=right:$u_3$] (m_3) at (1, 0) {};

\node[vertex, label=below:$s$] (t) at ($(m_2)!0.5!(m_3)-(0,1)$){};
\node[vertex, label=above:$t$] (s) at ($(j_1)!0.5!(j_2)+(0,1)$){};

\draw [reversepointedline, ultra thick, green, dashed] (j_1) -- node[edgelabel, very near start] {1} node[edgelabel, very near end] {2} (m_1);
\draw [reversepointedline, ultra thick, MyPurple] (j_1) -- node[edgelabel, very near start] {2} node[edgelabel, very near end] {1} (m_2);

\draw [reversepointedline, ultra thick, MyPurple] (j_2) -- node[edgelabel, very near start] {3} node[edgelabel, very near end] {1} (m_1);
\draw [dashed] (j_2) -- node[edgelabel, very near start] {2} node[edgelabel, very near end] {2} (m_2);
\draw [reversepointedline, ultra thick, green] (j_2) -- node[edgelabel, very near start] {1} node[edgelabel, very near end] {2} (m_3);

\draw [reversepointedline, ultra thick, green] (j_3) -- node[edgelabel, very near start] {1} node[edgelabel, very near end] {3} (m_2);
\draw [reversepointedline, ultra thick, MyPurple] (j_3) -- node[edgelabel, very near start] {2} node[edgelabel, very near end] {1} (m_3);

\draw [reversepointedline, gray] (j_2) [out=120,in=-60] -- node[edgelabel, near start] {1.5} (t);
\draw [reversepointedline, thick] (s) -- (j_1);
\draw [reversepointedline, thick] (s) -- (j_2);
\draw [reversepointedline, thick] (s) -- (j_3);
\draw [reversepointedline, gray] (s) -- node[edgelabel, near end] {1.5} (m_1);
\draw [reversepointedline, thick] (m_1) -- (t);
\draw [reversepointedline, thick] (m_2) -> (t);
\draw [reversepointedline, thick] (m_3) -> (t);

\end{tikzpicture}
\caption{The greedy algorithm fails to report the existence of a stable solution in this instance.}
\label{fig:wrong_order}
\end{figure}
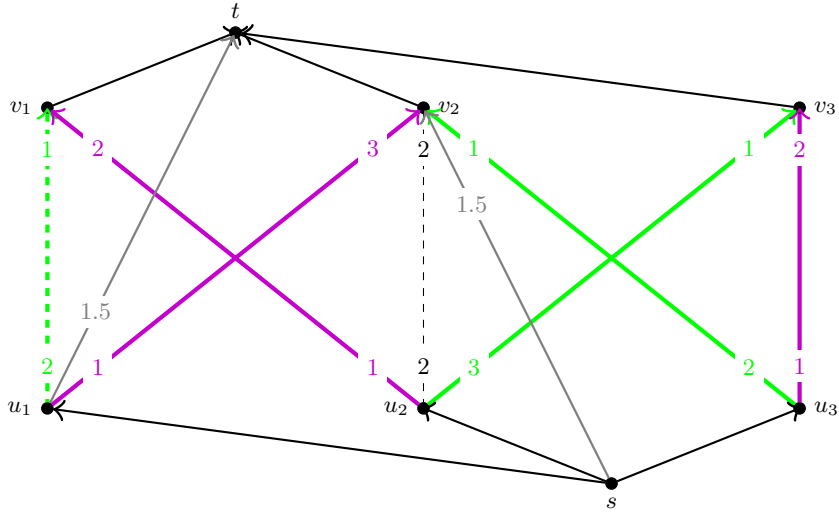

\begin{ex}[Stable flows with forbidden edges]
\label{sec:forbidden_fail}
In the unit-capacity network of Fig.~\ref{fig:wrong_order}, the dashed edges $u_1v_1$ and $u_2v_2$ form $P$, while the thin gray edges $sv_2$ and $u_1t$ are not part of the original graph but are added by the application of Lemma~\ref{le:single_forbidden}. The instance admits two stable flows. Both of them saturate all edges leaving $s$ and all edges entering~$t$. In the rest of the graph, stable flow $f_1$ is denoted by purple, and it sends one unit of flow along the edges in $\{ u_1v_2, u_2v_1, u_3v_3 \}$, while stable flow $f_2$ is denoted by green, and it sends one unit of flow along the edges in $f_2 =\{ u_1v_1, u_2v_3, u_3v_2 \}$. Since $u_1 v_1 \in P$ is used by $f_2$, only $f_1$ avoids~$P$. 
If tested separately, edge $u_2v_2$ fulfills both Properties~1 and~2 of Lemma~\ref{le:single_forbidden}, while $u_1v_1$ only fulfills Property~2. Yet requiring Property~1 for $u_2v_2$ and Property~2 for $u_1v_1$ by adding $sv_1$ and $u_2t$ to the graph (as the gray edges indicate) results in a graph where every stable flow uses both $sv_2$ and $u_1t$. This is because the only stable flow in the modified network with the edges $sv_2$ and $u_1t$ saturates edges $s u_1, s u_2, s u_3, s v_2, u_2 v_1, u_3 v_3, v_1 t, v_2 t, v_3 t$ and~$u_1 t$.
\end{ex}

We now sketch our algorithm that can deal with the presence of multiple forbidden edges. For any $A, B \subseteq E$, let us denote by $D[A|B]$ the network with vertices $V$ and edges $E \cup E^+(A) \cup E^-(B)$. We remind the reader that $E^+(A) := \{e^+ : e \in A\}$ and $E^-(B) := \{e^- : e \in B\}$. 
Our algorithm maintains a partition of the forbidden edges in two groups $P^+$ and $P^-$. Initially $P^+ = P$ and $P^- = \emptyset$. In every iteration, we compute a stable flow $f$ in $D[P^+|P^-]$. If $f(e^+) > 0$ for some $e \in P^+$, we move $e$ from $P^+$ to $P^-$ and repeat. If $f(e^+) = 0$ for all $e \in P^+$ but $f(e^-) > 0$ for some $e \in P^-$, we will show that no stable flow avoiding all forbidden edges exists in $D$. Finally, if we reach a flow $f$ where neither of these two things happens, then $f$'s restriction to $D$ is a stable flow in $D$ avoiding all forbidden edges, since $f(e^+) = 0$ or $f(e^-) = 0$ implies $f(e) = 0$ by choice of the ranks.

\begin{algorithm}[!h]
\caption{Stable flow with forbidden edges}
\label{alg:sf_forbidden}
Initialize $P^+ = P$ and $P^- = \emptyset$.\\
\Repeat{$f(e^+) = 0$ for all $e \in P^+$}{
	Compute a stable flow $f$ in $D[P^+|P^-]$.\\
  \If{$\exists\; e \in P^+$ with $f(e^+) > 0$}{
  	$P^+ := P^+ \setminus \{e\}$ and $P^- := P^- \cup \{e\}$
  }
}
    \If{$\exists\; e \in P^-$ with $f(e^-) > 0$}{
    	\Return $\emptyset$
	  }
    \Else{
    	\Return $f$
	  }
\end{algorithm}

Before proving its correctness, we present our algorithm run on the instance of Fig.~\ref{fig:wrong_order}.

\begin{ex}[Execution of Algorithm~\ref{alg:sf_forbidden}]
\label{sec:alg_forbidden}
Since $P = \left\{u_1v_1, u_2v_2 \right\}$, we initialize $P^{+}$ to be $\left\{u_1v_1, u_2v_2 \right\}$ and $P^{-}$ to be the empty set. This defines the network $D[P^+|P^-]$, which is $D$ complemented by $sv_1$ and $sv_2$. The stable flow $f$ computed by Algorithm~\ref{alg:poly_sf} in $D[P^+|P^-]$ saturates the edges $sv_1$, $v_1t$,  $su_2$, $u_2v_3$, $v_3t$, $su_3$, $u_3v_2$, and $v_2t$. Since $f(sv_1) >0$, the edge $u_1v_1$ is removed from $P^+$ and added to $P^-$.

In the second iteration, $D[P^+|P^-]$ is $D$ complemented by $u_1t$ and $sv_2$. The algorithm computes the stable flow in this network saturating the edges $su_1$, $u_1t$, $sv_2$, $v_2t$, $su_3$, $u_3v_3$, and $v_3t$. Because $f(sv_2) > 0$, the edge $u_2v_2$ is moved from $P^+$ to $P^-$.

In the third iteration,  $D[P^+|P^-]$ is $D$ complemented by $u_1t$ and $u_2t$. The algorithm computes the stable flow in this network saturating the edges $su_1$, $u_1v_2$, $v_2t$, $su_2$, $u_2v_1$, $v_1t$, $su_3$, $u_3v_3$, and $v_3t$. Since $P^+ = \emptyset$ and $f(e^-) = 0$ for all $e \in P^-$, the algorithm terminates by returning this flow.
\end{ex}

For the analysis of Algorithm~\ref{alg:sf_forbidden}, the following consequence of the augmenting path algorithm presented earlier (Algorithm~\ref{alg:poly_sf}) is helpful. It essentially states that removing an edge leaving $s$ and recomputing a stable flow cannot decrease the flow value on any other edge leaving~$s$. This observation will allow us to prove an important invariant of Algorithm~\ref{alg:sf_forbidden}.

\begin{lemma}
\label{le:add}
Let $f$ be a stable flow in $D$. Let $f'$ be a stable flow in $D' = D - e'$ for some edge $e' \in \delta^{+}(s)$. Then $f'(e) \geq f(e)$ for all $e \in \delta^{+}(s) \setminus \{e'\}$.
\end{lemma}
\proof
    We run Algorithm~\ref{alg:poly_sf} on the networks $D$ and $D'$, respectively, to obtain stable flows $f$ and $f'$. Recall that Algorithm~\ref{alg:poly_sf} uses an arbitrary but fixed order of the outgoing edges of $s$. We choose this order such that $e'$ comes last for the run in $D$. Observe that the algorithms run identically on both instances until $\propose[s]$ reaches $e'$ for the run on $D$ and terminates on $D'$, respectively. 
    Thus the flow $\bar{f}$ computed by the algorithm on $D$ right before $\propose[s]$ is advanced to $e'$ is identical to $f'$. 
    Further note that the algorithm does not increase the flow value on any edge $e \in \delta^+(s) \setminus \{e'\}$ after $\propose[s]$ has passed $e$, which comes before $e'$ by our choice of preferences. Hence $f(e) \leq \bar{f}(e) = f'(e)$.\qed

\begin{lemma}
\label{le:induction}
  Algorithm~\ref{alg:sf_forbidden} maintains the following invariant.
  There is a stable flow in $D$ avoiding $P$ if and only if there is a stable flow in $D[\emptyset|P^-]$ avoiding $P^+ \cup E^-(P^-)$.
\end{lemma}
\proof
Clearly, the invariant holds initially as $P^+ = P$ and $P^- = \emptyset$.
Now consider any later iteration of the algorithm in which $P^+, P^-$ are changed. Let $f_0$ be the computed stable flow in $D[P^+|P^-]$ and let $e_0$ be the edge with $f_0(e_0^+) > 0$ found in that iteration. 
Let $P^+_\text{old}, P^-_\text{old}$ and $P^+_\text{new}, P^-_\text{new}$ denote the partition before and after the update, i.e., $P^+_\text{new} = P^+_\text{old} \setminus \{e_0\}$ and $P^-_\text{new} = P^-_\text{old} \cup \{e_0\}$.

If there is a stable flow in $D[\emptyset|P^-_\text{new}]$ avoiding $P^+_\text{new} \cup E^-(P^-_\text{new})$, then this flow also avoids $P$, as for every $e \in P$ either $e \in P^+_\text{new}$ or $e^- \in E^-(P^-_\text{new})$ (note that in the latter case $e^-$ dominates $e$ at the start and ends at a terminal).

Conversely, if there is a stable flow in $D$ avoiding $P$, then by induction hypothesis there is a stable flow $f$ in $D[\emptyset|P^-_\text{old}]$ avoiding $P^+_\text{old} \cup E^-(P^-_\text{old})$. 
Note that $e^+_0$ starts at a terminal and recall that $f_0(e^+_0) > 0$ for the stable flow $f_0$ in $D[P^+_\text{old}|P^-_\text{old}]$.
By repeated application of Lemma~\ref{le:add}, deleting every $e^+ \in E^+(P^+_\text{old} \setminus \{e_0\})$ from $D[P^+_\text{old}|P^-_\text{old}]$, we obtain that $f'(e^+_0) > 0$ for every stable flow $f'$ in $D[\{e_0\}|P^-_\text{old}]$.
In particular, this means that Property 1 of Lemma~\ref{le:single_forbidden} fails for $f$ and $e_0$.
Therefore, by Lemma~\ref{le:single_forbidden}, Property 2 must hold for $f$, i.e., the extension of $f$ to $D[\emptyset|P^-_\text{old} \cup \{e_0^-\}] = D[\emptyset|P^-_\text{new}]$ with $f(e_0^-) = 0$ is a stable flow avoiding $P^+_\text{old} \cup E^-(P^-_\text{old}) \cup \{e_0^-\}$. As $P^+_\text{new} \subseteq P^+_\text{old}$ and $E^-(P^-_\text{new}) = E^-(P^-_\text{old}) \cup \{e^-_0\}$, this completes the induction.
\qed
\begin{lemma}
\label{le:no-instance}
  If Algorithm~\ref{alg:sf_forbidden} returns $\emptyset$, then no stable flow in $D$ avoids~$P$.
\end{lemma}
\proof
If the algorithm returns $\emptyset$, then the algorithm computed a stable flow $f$ in $D[P^+|P^-]$ with $f(e^+) = 0$ for all $e \in P^+$ but $f(e^-) > 0$ for some $e \in P^-$. Note that by Observation~\ref{ob:delete}, the restriction of $f$ is also stable in $D[\emptyset|P^-]$.
As $e^-$ is incident to a terminal, $f(e^-) > 0$ for every stable flow in $D[\emptyset|P^-]$. Therefore, by Lemma~\ref{le:induction}, there is no stable flow in $D$ avoiding~$P$.
\qed

\begin{lemma}
\label{le:yes-instance}
  If Algorithm~\ref{alg:sf_forbidden} returns flow $f$, then $f$ is stable in $D$ and it avoids~$P$.
\end{lemma}
\proof
If the algorithm returns flow $f$ then $f(e^+) = 0$ for all $e \in P^+$ and $f(e^-) = 0$ for all $e \in P^-$. Hence the restriction of $f$ to $E$ is stable and avoids $P^+ \cup P^- = P$.
\qed

The correctness of Algorithm~\ref{alg:sf_forbidden} follows immediately from the above lemmas. The running time of this algorithm is bounded by~$\mathcal{O}(|P| |E| \log |V|)$, as each stable flow $f$ can be computed in $\mathcal{O}(|E| \log |V|)$ time and in each round either $|P^+|$ decreases by one or the algorithm terminates.

\subsubsection{Stable matchings with forbidden edges}
\label{se:matchingforbidden}

Just as earlier, in Section~\ref{se:forcedmatching}, we finish this part with the direct interpretation of our results in the stable marriage instances. 

\begin{problem} \textsc{sm forbidden}\ \\
	\inp $\mathcal{I} = (G, r, P)$; a bipartite graph $G$, the preference ordering $r$ of vertices, and a set of forbidden edges~$P$. \\
	\ques Is there a stable matching avoiding all edges in~$P$?
\end{problem}

Let $A \cup B$ be the bipartition of the vertices.
One possibility to solve \textsc{sm forbidden} would be to transform it into an instance of \textsc{sf forbidden} by the standard transformation of bipartite matching to flow (directing all edges from $A$ to $B$ and augmenting the graph by a super source and a super sink connected to all vertices in $A$ and $B$, respectively). Running Algorithm~\ref{alg:sf_forbidden} on this instance gives a stable flow that can be transformed into a matching in the original instance. 

However, we can adapt the Algorithm~\ref{alg:sf_forbidden} to directly run on the matching instance as follows.
For forbidden each edge $e \in P$ we introduce a new vertex $v_e$. We maintain a partition of $P$ into sets $P_A$ and $P_B$, with $P_A = P$ and $P_B = \emptyset$ initially.
For each $e = ab \in P_A$ we introduce the edge $av_e$ to the graph with $r_a(av_e) = r_a(ab) - \varepsilon$, and for each edge $e = ab \in P_B$ we introduce the edge $bv_e$ instead with $r_{b}(bv_e) = r_{b}(ab) - \varepsilon$. We then compute a stable matching in the resulting graph. If an edge $av_e$ is in the matching for some $e \in P_A$ we remove $e$ from $P_A$ and add it to $P_B$. We then again compute a stable matching and repeat this procedure until no edge $av_e$ is in the matching for any $e = ab \in P_A$. 

If in the resulting matching the vertices $v_e$ for $e \in P$ are unmatched, i.e., also no edge $bv_e$ is used for any $e = ab \in P_B$, the matching is stable in the original graph and it does not use any edge in $P$ (due to the choice of the ranks). If not, using the same line of argumentation as in the proof of Lemma~\ref{le:induction} we can show that no stable matching avoiding $P$ exists. (Here, the bipartite structure of the graph yields a straightforward analogue of Lemma~\ref{le:add}. We remark that it is an open problem how to adapt this technique to the stable roommates problem for non-bipartite graphs.)


Our algorithm for several forbidden edges runs in $\mathcal{O}(|P| |E|)$ time, because computing stable matchings in each of the at most $|P|$ rounds takes only $\mathcal{O}(|E|)$ time. 
With this running time, it is somewhat slower than the best known methods~\cite{DFFS03,FIM07} that require only $\mathcal{O}(|E|)$ time, but it is a reasonable assumption that the number of forbidden edges is small.

\subsection{Forced and forbidden edges}

If both forced and forbidden edges occur in the same instance, then they can be handled by our two algorithms, applying them one after the other. First, all forced edges in the graph $D$ are substituted by the construction discussed in Section~\ref{se:multipleforced}, obtaining the graph $D_Q$ where the edges in $Q$ are replaced by artificial edges $\bar{Q}$. The following corollary is a direct implication of Theorem~\ref{thm:multiple-forced-edges}.

\begin{corollary}
\label{cor:forced-and-forbidden}
There is a stable flow in $D$ saturating all edges in $Q$ and avoiding all edges in $P$ if and only if there is a stable flow in $D_Q$ saturating all edges in $\bar{Q}$ and avoiding all edges in $P$.
\end{corollary}

We now run Algorithm~\ref{alg:sf_forbidden} from Section~\ref{se:multipleforbidden} on $D_Q$.
If the algorithm asserts that no stable flow in $D_Q$ avoiding $P$ exists, then by Corollary~\ref{cor:forced-and-forbidden}, there is no stable flow in $D$ saturating all edges in $Q$ and avoiding all edges in $P$.
If, instead, the algorithm returns a stable flow $f'$ avoiding $P$, we check whether it also saturates all edges in $\bar{Q}$. If this is the case, the corresponding flow in $D$ is a stable flow avoiding $P$ and saturating all edges in $Q$. If there is an edges $e \in \bar{Q}$ with $f'(e) < c(e)$, then this is true for every stable flow in $D_Q$ by the Rural Hospital Theorem (Theorem~\ref{same_flow_value}) and hence, no flow saturating all edges in $Q$ exists in~$D$.

The procedure described above runs in time $\mathcal{O}(|P| |E| \log{|V|})$, as $D_Q$ can be constructed in time linear in $|E|$ and the number of edges and vertices in $D_Q$ is at most twice the number of edges and vertices in $D$, respectively (remember that we already argued in Remark~\ref{rem:sf_forbidden_reduction} that the initial transformation of the instance in Section~\ref{sec:sf_forbidden_reduction} does not change this asymptotic running time).
We conclude the following result:

\begin{theorem}
	\textsc{sf restricted} can be solved in $\mathcal{O}(|P| |E| \log{|V|})$ time.
\end{theorem}

\section{Stable multicommodity flows}
\label{sec:smcf}
In this section we turn our attention to stable multicommodity flows. 
We first present the original definition of this concept by Kir\'aly and Pap~\cite{KP13a} and outline their results, including the existence of a stable solution. We then proceed to our results: a reduction of the general model to a much simpler special case and a hardness proof for deciding the existence of an integral solution.

\subsection{Problem definition}

Multicommodity networks model scenarios in which a common network is used by several commodities. For example, roads serve personal vehicles, and also various sorts of commercial transport vehicles. While each person and each type of goods has its own origin and destination, they all share the same roads, which have a capacity on all vehicles altogether and sometimes also separately on a specific type of vehicle.

A \emph{multicommodity network} $(D, c^i, c), 1 \leq i \leq n$ consists of a directed graph $D = (V, E)$, non-negative commodity capacity functions $c^i: E\rightarrow\mathbb{R}_{\geq 0}$ for all the $n$ commodities and a non-negative cumulative capacity function $c:E\rightarrow\mathbb{R}_{\geq 0}$ on~$E$. For every commodity $i$, there is a \emph{source} $s^i \in V$ and a \emph{sink} $t^i \in V$, also referred to as the \emph{terminals of commodity~$i$}.

\begin{definition}[multicommodity flow]
	\label{def:mf}
	A set of functions $f^i:E\rightarrow\mathbb{R}_{\geq 0}$, $1 \leq i \leq n$ is a \emph{multicommodity flow} \index{multicommodity flow} if it fulfills all of the following requirements:
	\begin{enumerate}
		\item capacity constraints for commodities:\\ $f^i(uv) \leq c^i(uv)$ for all $uv \in E$ and commodity~$i$;
		\item cumulative capacity constraints:\\ $f(uv) = \sum_{1 \leq i \leq n}{f^i(uv)} \leq c(uv)$ for all $uv \in E$;
		\item flow conservation:\\ $\sum_{uv \in E}{f^i(uv)} = \sum_{vw \in E}{f^i(vw)}$ for all $i: 1 \leq i \leq n$ and $v \in V \setminus \{s^i, t^i\}$.
	\end{enumerate}
\end{definition}

The concept of stability was extended to multicommodity flows by Kir\'aly and Pap~\cite{KP13a}. A stable multicommodity flow instance $\mathcal{I} = (D, c^i, c, r_E, r_V^i), 1 \leq i \leq n$ comprises a network $(D, c^i, c), 1 \leq i \leq n$, \emph{edge preferences} $r_E$ over commodities, and \emph{vertex preferences} $r_V^i, 1 \leq i \leq n$ over incident edges for commodity~$i$. Each edge $uv$ ranks all commodities in a strict order of preference. Separately for every commodity~$i$, each non-terminal vertex ranks its incoming and also its outgoing edges strictly with respect to commodity~$i$. Note that these preference orderings of $v$ can be different for different commodities and they do not depend on the edge preferences $r_E$ over the commodities. If edge $uv$ prefers commodity $i$ to commodity $j$, then we write~$\rank_{uv} (i) < \rank_{uv} (j)$. Analogously, if vertex $v$ prefers edge $vw$ to~$vz$ with respect to commodity~$i$, then we write~$\rank_{v}^{i} (vw) < \rank_{v}^{i} (vz)$. We denote the flow value with respect to commodity $i$ by $f^i = \sum_{u \in V}{f^i(s^iu)}$.

\begin{definition}[stable multicommodity flow]
\label{def:mcblocking}
		A \emph{blocking walk with respect to commodity $i$} of a multicommodity flow $f$ is a directed walk $W=\langle v_{1}, v_{2}, ..., v_{k} \rangle$ such that all of the following properties hold:
		\begin{enumerate}
			\item $f^i(v_jv_{j+1}) < c^i(v_jv_{j+1})$	for each edge $v_jv_{j+1}$, $j=1, ...,k-1$;
			\item $v_1 = s^i$ or there is an edge $v_{1}u$ such that $f^i(v_{1}u) > 0$ and $\rank_{v_{1}}^i (v_{1} v_{2}) < \rank_{v_{1}}^i (v_{1}u)$; 
			\item $v_{k} = t^i$ or there is an edge $wv_{k}$ such that $f^i( wv_{k})>0$ and $\rank_{v_{k}}^i (v_{k-1} v_{k}) < \rank_{v_{k}}^i (wv_{k})$;
			\item if $f(v_j v_{j+1}) = c(v_j v_{j+1})$, then there is a commodity $i'$ such that $f^{i'}(v_j v_{j+1}) > 0$ and $\rank_{v_j v_{j+1}} (i) < \rank_{v_j v_{j+1}} (i')$.
		\end{enumerate}
		A multicommodity flow is \emph{stable}, if there is no blocking walk with respect to any commodity.
\end{definition}

In words, a walk blocks the multicommodity flow with respect to commodity $i$ if both the starting and end vertices of the walk are willing to reroute some units of flow of commodity $i$ along it, moreover, the edges along the walk either have free capacity for forwarding these or they are inclined to drop some units of flow of another commodity. This last point can be seen as a clear difference to single-commodity stable flows. Due to point~4, Definition~\ref{def:mcblocking} allows saturated edges to occur in a blocking walk with respect to commodity $i$, provided that these edges are inclined to trade in some of their forwarded commodities for more flow of commodity~$i$. On the other hand, the role of edge preferences is limited: blocking walks still must start at vertices who are willing to reroute or send extra flow along the first edge of the walk according to their vertex preferences with respect to commodity~$i$.

\begin{pr} \textsc{smf}\ \\
\label{def:smf}
	\inp $\mathcal{I} = (D, c^i, c, r_E, r_V^i)$, $1 \leq i \leq n$ ; a directed multicommodity network $(D,c^i, c)$, $1 \leq i \leq n$, edge preferences over commodities $r_E$ and vertex preferences over incident edges $r_V^i, 1 \leq i \leq n$. \\
	\ques Is there a stable multicommodity flow?
\end{pr}

\begin{theorem}[Kir\'aly, Pap~\cite{KP13a}]
	\label{th:ppad}
	A stable multicommodity flow exists for any instance, but it is $\PPAD$-hard to find.
\end{theorem}

Kir\'aly and Pap use a polyhedral version of Sperner's lemma~\cite{KP09b} to prove the existence result.
$\PPAD$-hardness~\cite{Pap94} is considered a somewhat weaker evidence of intractability than $\NP$-hardness that applies for problems whose decision versions have a 'yes' answer for sure. Note that \textsc{smf} is one of the very few problems in stability~\cite{BKPW15} where a stable solution exists, but no extension of the Gale-Shapley algorithm is known to solve it -- not even a variant with exponential running~time.

\subsection{Problem simplification}

The definition of \textsc{smf} involves many distinct components and constraints. It is natural to investigate how far the model can be simplified without losing any of its generality. In particular, Kir\'{a}ly and Pap~\cite{KP13a} pose an open question on the $\PPAD$-hardness of the problem if there are no individual capacities. Here we give a positive answer to this and further intuitive questions on possible restricted cases. It turns out that the majority of the commodity-specific input data can be dropped, as shown by Theorem~\ref{th:smf_simple}.  This result not only simplifies the instance, but it also sheds light to the most important characteristic of the problem, which seems to be the preference ordering of edges over commodities.

\begin{theorem}
	There is a polynomial-time transformation that, given an instance $\mathcal{I}$ of \textsc{smf}, constructs an instance $\mathcal{I}'$ of \textsc{smf} with the following properties:
    \begin{enumerate}
      \item all commodities have the same source and sink,
      \item at each vertex, the preference lists are identical for all commodities,
      \item there are no commodity-specific edge capacities,
    \end{enumerate}
    and there is a polynomially computable bijection between the stable multicommodity flows of $\mathcal{I}$ and the stable multicommodity flows of $\mathcal{I}'$. The bijection preserves integrality.
    \label{th:smf_simple}
\end{theorem}

\begin{proof}
We present the construction in three steps, each ensuring one of the properties without destroying those established before.

\begin{enumerate}
      \item \emph{All commodities have the same source and sink.\\}
We introduce two new super terminals $s^*$ and~$t^*$. These will substitute all commodity-specific sources and sinks. For every commodity $i$ and its terminals $s^i$ and $t^i$, we introduce the edges $s^*s^i$ and $t^it^*$ with capacities $c^i(s^*s^i) = c(s^*s^i) = \sum_{e \in \delta^+(s^i)} c(e)$ and $c^i(t^it^*) = c(t^it^*) = \sum_{e \in \delta^-(t^i)} c(e)$. These edges cannot carry any other commodity: $c^{j}(s^{*}s^{i})=c^{j}(t^{i}t^{*})=0$ for  all $j \neq i$. We assign arbitrary ranks to the edges originally incident to $s^i$ or $t^i$ and put $s^*s^i$ and $t^it^*$ to the end of the preference list of $s^i$ and $t^i$ for all commodities. Finally, we set $s^*$ and $t^*$ as source and sink for every commodity $i$.
It is easy to verify that a flow $f$ is stable in the original network $D$ if and only if the natural extension of $f$ to the added edges is a stable flow.
 \smallskip
\item \emph{At each vertex, the preference lists over the edges are identical for all commodities.\\}
      The main idea here is to substitute every edge by a gadget that separates different commodities. Then the edges can be ranked in a single preference list, since each edge is designated to carry its own commodity only and for edges carrying a specific commodity, the list on other edges is irrelevant. 
			
		For any $e \in E$, we remove $e = uv$ from the graph and replace it by the construction shown in Fig.~\ref{fi:identical_pref}. We introduce two new vertices $v'_e$ and $v''_e$ and add the edge $v'_{e}v''_{e}$ with $c(v'_{e}v''_{e}) = c^i(v'_{e}v''_{e}) = c(e)$ for every commodity $i$. We also add $n$ new edges $e'_i$ for $1 \leq i \leq n$ from $u$ to $v'_{e}$. We set $c(e'_i) = c^i(e'_i) = c^i(e)$, $c^{j}(e'_i) = 0$ for $j \neq i$, and $\rank_{u}(e'_i) = |E|i + \rank^i_{u}(e)$. We choose $\rank_{v'_e}(e'_i)$ arbitrarily.
Likewise, we add $n$ new edges $e''_i$ for $1 \leq i \leq n$ from $v''_{e}$ to $v$. We set $c(e''_i) = c^i(e''_i) = c^i(e)$, $c^{j}(e''_i) = 0$ for $j \neq i$, and $\rank_{v}(e''_i) = |E|i + \rank^i_{v}(e)$. We choose $\rank_{v''_e}(e'_i)$ arbitrarily. Let $D'$ be the network resulting from this modification.

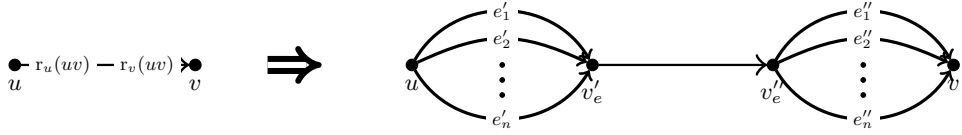
\begin{figure}[t]
\centering
\begin{tikzpicture}[scale=1, transform shape]
\pgfmathsetmacro{\d}{0.8}
\pgfmathsetmacro{\b}{2.4}

\node[vertex, label=below:$u$] (u) at (0, 0) {};
\node[vertex, label=below:$v'_e$] (v') at (\b, 0) {};
\node[vertex, label=below:$v''_e$] (v'') at (2*\b, 0) {};
\node[vertex, label=below:$v$] (v) at (3*\b, 0) {};
\node[vertex, label=below:$u$] (u0) at (-\b*2.2, 0) {};
\node[vertex, label=below:$v$] (v0) at (-\b*1.2, 0) {};

\draw [->, very thick] (u) to[out=60,in=120, distance=1cm] node[edgelabel, scale=\d] {$e'_1$} (v');
\draw [->, very thick] (u) to[out=25,in=155, distance=1cm] node[edgelabel, scale=\d] {$e'_2$} (v');
\draw [->, very thick] (u) to[out=-60,in=-120, distance=1cm] node[edgelabel, scale=\d]  {$e'_n$} (v');
\node[vertex, fill=black, scale=0.4] (a1) at ($(u)!0.5!(v')$) {};
\node[vertex, fill=black, scale=0.4] (a2) at ($(u)!0.5!(v')-(0, 0.2)$) {};
\node[vertex, fill=black, scale=0.4] (a3) at ($(u)!0.5!(v')-(0, 0.4)$) {};

\draw [pointedline] (v') -- (v'');
\draw [->, very thick] (v'') to[out=60,in=120, distance=1cm] node[edgelabel, scale=\d] {$e''_1$} (v);
\draw [->, very thick] (v'') to[out=25,in=155, distance=1cm] node[edgelabel, scale=\d] {$e''_2$} (v);
\draw [->, very thick] (v'') to[out=-60,in=-120, distance=1cm] node[edgelabel, scale=\d]  {$e''_n$} (v);
\node[vertex, fill=black, scale=0.4] (a1) at ($(v)!0.5!(v'')$) {};
\node[vertex, fill=black, scale=0.4] (a2) at ($(v)!0.5!(v'')-(0, 0.2)$) {};
\node[vertex, fill=black, scale=0.4] (a3) at ($(v)!0.5!(v'')-(0, 0.4)$) {};

\draw [pointedline] (u0) -- node[edgelabel, near start, scale=\d] {$\rank_u(uv)$} node[edgelabel, near end, scale=\d] {$\rank_v(uv)$} (v0);

\draw [arrow](-\b*0.8,0)-- (-\b*0.5,0);
\end{tikzpicture}
\caption{The gadget ensuring that the preference lists of each vertex are identical for all commodities.}
\label{fi:identical_pref}
\end{figure}

If $f$ is a stable flow in $D$, then we define a flow $f'$ in $D'$ as follows. For every commodity $i$ and every $e \in E$, we set $f'^i(e'_i) = f'^i(v'_{e}v''_{e}) = f'^i(e''_i) = f^i(e)$ and we set $f'^j(e'_i) = f'^j(e''_i) = 0$ for $j \neq i$. It is easy to check that $f'$ is a stable flow in $D'$ and that the mapping from $f$ to $f'$ is a bijection between stable flows in $D$ and $D'$.
\smallskip 

\item \emph{There are no commodity-specific capacities.\\}
Finally we ensure that $c^i(e) = c(e)$ for all $i$ and all $e \in E$, which implies that the commodity-specific capacities do not play any role. To this end, we introduce a new commodity $i^*$. Each edge will be replaced by a gadget in which the capacity on a specific commodity translates into an edge willing to carry $i^*$ rather than forwarding more flow of the specific commodity.

Note that the transformation described in point 2 above already ensures that for every edge $e \in E$ one of the following is true: Either $c^{i}(e) = c(e)$ for all $i$, or there is an $i$ such that $c^i(e) = c(e)$ and $c^j(e) = 0$ for all $j \neq i$. We only have to deal with the latter case, that is, edge $e$ being designated to carry commodity $i$ only, up to its full capacity. Let edge $e$ and commodity $i$ be such a pair.

We replace $e = uv$ by the gadget $H_{e,i}$, depicted in Fig.~\ref{fi:comm}. First, four new vertices $u', u'', v'$ and $v''$ are introduced. We add the edges $uu'$, $u'v'$, $v'v$, $su''$, $u''v''$, $v''t$, $u''u'$ and $v'v''$, all with capacity $c(e)$. 
For the edges $su''$, $u''v''$, $v''t$, $u''u'$ and $v'v''$ the new commodity $i^*$ is on top of their preference list, followed by all other commodities in arbitrary order. For edge $u'v'$ commodity $i$ is first on the list, $i^*$ is second, followed by all other commodities in arbitrary order. For the edges $uu'$ and $v'v$,  commodity $i^*$ is last on the list, the rank of the other commodities is arbitrary.
For the vertex preferences, we set $\rank_{u''}(u''u') < \rank_{u''}(u''v'')$ and 
$\rank_{v''}(v'v'') < \rank_{v''}(u''v'')$, as well as $\rank_{u'}(u''u') < \rank_{u'}(uu')$ and $\rank_{v'}(v'v'') < \rank_{v'}(v'v)$. We further set $\rank_{u}(uu') = \rank_{u}(e)$ and $\rank_{v}(v'v) = \rank_{v}(e)$.
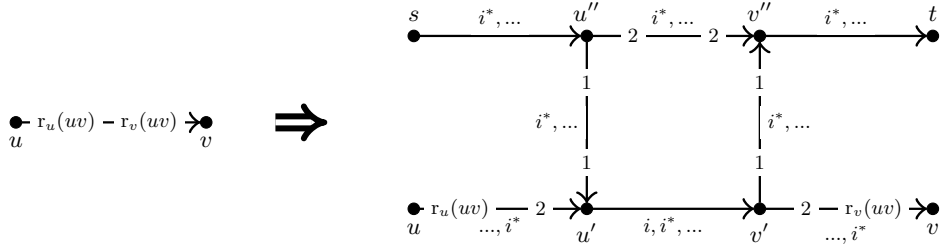
\begin{figure}[t]
\centering
\begin{tikzpicture}[scale=1, transform shape]
\pgfmathsetmacro{\d}{0.85}
\pgfmathsetmacro{\b}{2.3}

\node[vertex, label=below:$u'$] (u') at (0, 0) {};
\node[vertex, label=above:$u''$] (u'') at (0, \b) {};
\node[vertex, label=below:$u$] (u) at (-\b, 0) {};
\node[vertex, label=below:$v$] (v) at (2*\b, 0) {};
\node[vertex, label=below:$v'$] (v') at (\b, 0) {};
\node[vertex, label=above:$v''$] (v'') at (\b, \b) {};

\node[vertex, label=above:$s$] (s) at (-\b, \b) {};
\node[vertex, label=above:$t$] (t) at (2*\b, \b) {};

\node[vertex, label=below:$u$] (u0) at (-\b*3.3, 0.5*\b) {};
\node[vertex, label=below:$v$] (v0) at (-\b*2.2, 0.5*\b) {};

\draw [pointedline] (u) -- node[edgelabel, below, scale=\d] {$..., i^*$} node[edgelabel, near start, scale=\d] {$\rank_u(uv)$} node[edgelabel, near end, scale=\d] {2} (u');
\draw [pointedline] (u') -- node[edgelabel, below, scale=\d] {$i, i^*, ...$} (v');
\draw [pointedline] (v') -- node[edgelabel, scale=\d, yshift=-4mm] {$..., i^*$} node[edgelabel, near start, scale=\d] {2} node[edgelabel, near end, , scale=\d, xshift=-2mm] {$\rank_v(uv)$} (v);
\draw [pointedline] (u'') -- node[edgelabel, left, scale=\d] {$i^*, ...$} node[edgelabel, near start, scale=\d] {1} node[edgelabel, near end, scale=\d] {1} (u');
\draw [pointedline] (v') -- node[edgelabel, right, scale=\d] {$i^*, ...$} node[edgelabel, near start, scale=\d] {1} node[edgelabel, near end, scale=\d] {1} (v'');
\draw [pointedline] (s) -- node[edgelabel, above, scale=\d] {$i^*, ...$} (u'');
\draw [pointedline] (u'') -- node[edgelabel, above, scale=\d] {$i^*, ...$} node[edgelabel, near start, scale=\d] {2} node[edgelabel, near end, scale=\d] {2} (v'');
\draw [pointedline] (v'') -- node[edgelabel, above, scale=\d] {$i^*, ...$} (t);

\draw [pointedline] (u0) -- node[edgelabel, near start, scale=\d] {$\rank_u(uv)$} node[edgelabel, near end, scale=\d, xshift=-1mm] {$\rank_v(uv)$} (v0);

\draw [arrow](-\b*1.8,\b/2)-- (-\b*1.5,\b/2);

\end{tikzpicture}
\caption{The gadget ensuring that there are no commodity-specific capacities.}
\label{fi:comm}
\end{figure}

Let us denote the modified network by $\bar{D}$. 
For a stable flow $f$ in the original network $D$, we define a flow $\bar{f}$ in $\bar{D}$ as follows. 
For edges $e$ that were not replaced by a gadget in $\bar{D}$, we set $\bar{f}^i(e) = f^i(e)$ for all $i$.
For every $e$ that was replaced by a gadget (because $c^i(e) = c(e)$ and $c^j(e) = 0$ for all $j \neq i$), we set the flow values within the gadget as follows.
For the new commodity $i^*$ we set $\bar{f}^i(uu') = \bar{f}^i(u'v') = \bar{f}^i(v'v) = f^i(e)$, and we set $\bar{f}^{i*}(u''u') = \bar{f}^{i*}(u'v') = \bar{f}^{i*}(v'v'') = c(e) - f^i(e)$, so that $u'v'$ is saturated with its two top-ranked commodities. Furthermore we set $\bar{f}^{i*}(su'') = \bar{f}^{i*}(v''t) = c(e)$, and $\bar{f}^{i*}(u''v'') = f^i(e)$.
All other flow values are set to zero within the gadget (recall that $f^j(e) = 0$ for all $j \neq i$).
\begin{claim}
The flow $\bar{f}$ is stable in $\bar{D}$.
\end{claim}
\begin{proof}
We have constructed $\bar{f}$ so that it respects all capacities and fulfills flow conservation in $\bar{D}$.
To see that $\bar{f}$ is a stable flow, assume by contradiction that there is an $\bar{f}$-blocking walk $\bar{W}$ for some commodity~$j$.


First assume $\bar{W}$ starts in the interior of a gadget, i.e., with an edge of a gadget $H_{e,i}$ different from $uu'$. We eliminate the edges of the gadget one by one to show that this is not possible.
\begin{itemize}
	\item $\bar{W}$ cannot start with $su''$, as this edge is saturated with its most preferred commodity~$i^*$. 
	\item $\bar{W}$ also cannot start with $u''v''$, $u'v'$, or $v'v$, as these edges are the last-choice outgoing edges on the preference lists of $u''$, $u'$ and $v'$ respectively.
	\item If $\bar{W}$ starts at $u''u'$, then $j = i^*$, because this is the only commodity on the dominated edge $u''v''$. But then $\bar{W}$ must end at $u'$ because $u'v'$ is saturated with commodities it ranks at least as high as $i^*$. However, $\bar{f}^{i^*}(uu') = 0$, so $\bar{W}$ does not dominate $f$ at $u'$. 
	\item Finally, if $\bar{W}$ starts with $v'v''$, then $j \neq i^*$ because $\bar{f}^{i^*}(v'v) = 0$. But it can neither end at $v''$ as $v''$ only receives commodity $i^*$ from $u''v''$, nor can it continue as $v''t$ is saturated with its favorite commodity.
	\end{itemize}
	
	We conclude that $\bar{W}$ cannot start in the interior of a gadget. By a symmetric argument, $\bar{W}$ cannot end in the interior of a gadget, i.e., with an edge of a gadget $H_{e,i}$ different from $v'v$.
\smallskip

Thus, if $\bar{W}$ contains any edge of a gadget $H_{e,i}$, it must traverse all the edges $uu', u'v', v'v$ of the gadget.
As $u'v'$ is saturated with commodities $i$ and $i^*$, we conclude that $j = i$ and $c(e) - f^{i}(e) = \bar{f}^{i^*}(u'v') > 0$. We replace all such segments $uu', u'v', v'v$ from any traversed gadget $H_{e, i}$ with the corresponding edge $e$ and get a walk $W$ in $D$. Because $f^{i}(e) < c(e)$ for all inserted edges, $W$ is a blocking walk for $f$, contradicting the stability of~$f$.~\myqed
\end{proof}

It is easy to see that the mapping defined by $\phi(f) = \bar{f}$ is injective, and as argued above, preserves stability. We now show that it is indeed a bijection from stable flows in $D$ to stable flows in $\bar{D}$.

\begin{claim}
For any stable flow $y$ in $\bar{D}$, there is a stable flow $f$ in $D$ with $\phi(f) = y$.
\end{claim}
\begin{proof}
Let $y$ be a stable flow in $\bar{D}$. 
Consider a gadget $H_{e,i}$. By contradiction assume $y^{i^*}(uu') > 0$. Then $y^{i^*}(su'') = y^{i^*}(u''u') = c(e)$ as otherwise either $\langle s, u'', u' \rangle$ or $\langle u'', u' \rangle$ is a blocking walk for commodity $i^*$. But then $y^{i^*}(uu') + y^{i^*}(u''u') > c(e) \geq y^{i^*}(u'v')$, contradicting flow conservation. Hence $y^{i^*}(uu') = 0$ and, by a symmetric argument, $y^{i^*}(v'v) = 0$. As no flow of commodity $i^*$ enters or leaves $H_{e,i}$, and the path $\langle s, u'', v'', t\rangle$ is not blocking, we conclude that $y^{i^*}(su'') = y^{i^*}(v''t) = c(e)$. By flow conservation, $y^{i^*}(u''u') = y^{i^*}(u'v') = y^{i^*}(v'v'') = c(e) - y^{i^*}(u''v'')$.
Since the path $\langle u'', u', v', v''\rangle$ is not blocking and $i$ is the only commodity that comes before $i^*$ on an edge of that path, we conclude that $y^{i^*}(u'v') + y^{i}(u'v') = c(e)$. Hence, by flow conservation, $y^{i}(uu') = y^{i}(u'v') = y^{i}(v'v) = c(e) - y^{i^*}(v'v'')$, and $y^{j}(e') = 0$ for all $j \notin \{i, i^*\}$ and all edges $e'$ in the gadget $H_{e,i}$.

Now define $f$ by setting $f^i(e) = y^i(u'v')$ for every gadget $H_{e,i}$ in $\bar{D}$ and $f^i(e) = y^i(e)$ for all edges in $E \cap E_{\bar{D}}$ and all commodities $i$. Using the above observations, it is easy to check that $\phi(f) = y$ and that $f$ fulfills flow conservation and respects all capacity constraints (in particular $f^j(e) = y^j(u'v') = 0$ for all $j \neq i$ at any gadget $H_{e,i}$). To see that $f$ is a stable flow, assume by contradiction that there is a blocking walk $W$ for $f$ and commodity $i$. We obtain a walk $\bar{W}$ in $\bar{D}$ by replacing the edges of $W$ with the corresponding gadgets $H_{e,i}$. At any such edge, $f^{i}(e) < c(e)$ because $W$ is blocking with respect to $i$ and $i$ is the only commodity that can traverse $e$. Hence, $y^{i}(uu') = y^{i}(u'v') = y^{i}(v'v) < c(e)$. Also, as the preference lists of non-gadget vertices are the same in $D$ and $\bar{D}$, $\bar{W}$ is indeed a blocking walk for $y$ contradicting its stability. \myqed
\end{proof}
\end{enumerate}
It is easy to check that all transformations described above can be carried out in polynomial time and that integral stable flows in the original graph correspond to integral stable flows in the transformed graph.\qed
\end{proof}

\subsection{Integral multicommodity stable flows}

First we modify Definition~\ref{def:smf} so that it describes the integral version of \textsc{smf}. Then we carefully analyze an example network with no integral solution. This network is used in the last part of this subsection, in which we present our hardness proof.

\begin{pr} \textsc{ismf}\ \\
\label{def:ismf}
	\inp $\mathcal{I} = (D, c^i, c, r_E, r_V^i)$, $1 \leq i \leq n$ ; a directed multicommodity network $(D,c^i, c)$, $1 \leq i \leq n$, edge preferences over commodities $r_E$ and vertex preferences over incident edges $r_V^i, 1 \leq i \leq n$. \\
	\ques Is there a stable multicommodity flow with integral $f^i(uv)$ values for all $uv \in E$ and $1 \leq i \leq n$?
\end{pr}

Kir\'aly and Pap~\cite{KP13a} give,  for every integer~$N$, an example instance with $N$ commodities and $N$ vertices, where no stable multicommodity flow exists with denominators at most~$N$. Here we present a small and slightly modified version of that instance as an example and later use it as a gadget in our hardness proof.

\begin{figure}[t]
	\centering
		\begin{tikzpicture}[scale=1, transform shape,->]
		\pgfmathsetmacro{\b}{3}
		\pgfmathsetmacro{\d}{3}
		\pgfmathsetmacro{\a}{90}
		\pgfmathsetmacro{\c}{45}
		
		\node[vertex, fill=white, label=above:$u$] (u) at (0, 0) {};
		\node[vertex, label=below:$v_1$] (v1) at (-\b, -2*\d) {};
		\node[vertex, label={below, yshift=-2mm}:$v_2$] (v2) at (0, -\d) {};
		\node[vertex, label=below:$v_3$] (v3) at (\b, -2*\d) {};
		
		\draw [pointedline2] (u) to[out=180+\c,in=\a] node[edgelabel, very near end] {2} node[edgelabel, text=MyPurple] {1} (v1);
		\draw [pointedline2] (u) to[out=180+\c,in=180-\c] node[edgelabel, near end] {2} node[edgelabel, text=MyPurple] {2} (v2);
		\draw [pointedline2] (u) to[out=-\c,in=180-\a] node[edgelabel, very near end] {2} node[edgelabel, text=MyPurple] {3} (v3);
		
		\draw [pointedline2] (v1) to[out=180-\c,in=180] node[edgelabel, very near start] {2} node[edgelabel, text=MyPurple] {2} (u);
		\draw [pointedline2] (v2) to[out=\c,in=-\c] node[edgelabel, near start] {2} node[edgelabel, text=MyPurple] {3} (u);
		\draw [pointedline2] (v3) to[out=\c,in=0] node[edgelabel, very near start] {2}node[edgelabel, text=MyPurple] {1} (u);
		
		\draw [pointedline, MyPurple] (v1) -- node[edgelabel, near start, text=black] {1} node[edgelabel, near end, text=black] {1} node[edgelabel] {3,1} (v2);
		\draw [pointedline, MyPurple] (v2) -- node[edgelabel,  near start, text=black] {1} node[edgelabel, near end, text=black] {1} node[edgelabel] {1,2} (v3);
		\draw [pointedline, MyPurple] (v3) -- node[edgelabel,  near start, text=black] {1} node[edgelabel, near end, text=black] {1} node[edgelabel] {2,3} (v1);
		
		\end{tikzpicture}
\caption{The edge preferences are marked with colored labels in the middle of edges, while $r_V^i$ is black and closer to the vertices. For all edges, $c = 1$. The purple edges of the triangle can forward two commodities, while the bent black edges can carry only one commodity.}
\label{fi:no_ismf}
\end{figure}
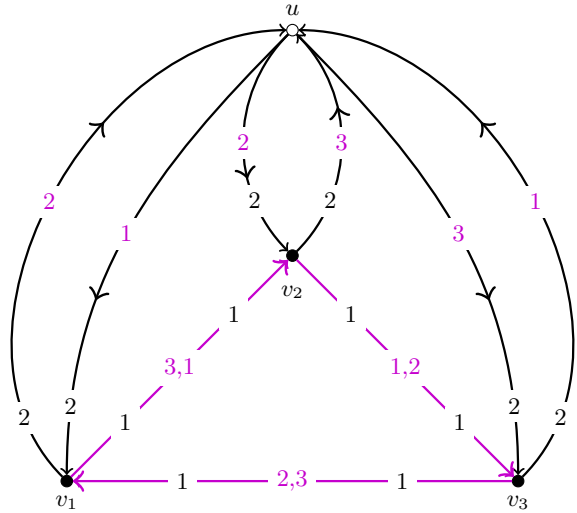

\begin{ex}[ISMF instance with no solution]
\label{ex:triangle}
Consider the network depicted in Fig.~\ref{fi:no_ismf}.
We consider two variants of an \textsc{ismf} instance in this network. In both cases, $u$ is the only terminal vertex in the graph, but the variants differ in that either $3$ or only $2$ commodities are present:
\begin{enumerate}
	\item $s^1= s^2= s^3= t^1 = t^2 = t^3 = u$  (see Lemma~\ref{le:noint}) and
	\item $\exists i \in \{ 1,2,3\}: \{s^i, t^i\} = \emptyset$ (see Lemma~\ref{le:int}). 
\end{enumerate}
We will show below that in the first case, the instance admits no integer multicommodity flow, whereas such a flow exists in the second case.

The edge capacities with respect to commodities are 1 for the commodities that appear in $r_E$ for the specific edge and 0 for the remaining commodities. All edges have cumulative capacity~1. The vertex preferences are the same for all commodities: $v_1, v_2$ and $v_3$ are inclined to receive and send the flow along the edges between themselves rather than trading with~$u$. Each commodity $i$ has a unique feasible cycle $C^i$ through $u$ and it is easy to see that due to the choice of the $c^i$ functions, no other cycle or terminal-terminal path exists in the network.
\begin{itemize}
	\item $C^1 = \langle u, v_1, v_2, v_3, u \rangle$
	\item $C^2 = \langle u, v_2, v_3, v_1, u \rangle$
	\item $C^3 = \langle u, v_3, v_1, v_2, u \rangle$
\end{itemize}
\end{ex}

\begin{lemma}
\label{le:noint}
	If $s^1= s^2= s^3= t^1 = t^2 = t^3 = u$, then there is no integer stable multicommodity flow.
\end{lemma}
\proof

Assume that there is an integral stable multicommodity flow $f$ in the instance. The empty flow cannot be $f$, because there is a cycle running through $u$ for each commodity and such cycles block the empty flow. Without loss of generality we can now assume that $C^1$ is saturated by commodity~1:
$$f^1(uv_1) = f^1(v_1v_2) = f^1(v_2v_3) = f^1(v_3u) =1,$$
while all other flow values must be 0 due to commodity capacity constraints on edges. This flow is blocked by commodity~3 on the cycle $\langle u, v_3, v_1, v_2, u \rangle$. It is easy to see that analogous arguments work for $C^2$ and $C^3$ as well. Thus, no integer stable flow exists in the graph.
\qed

\begin{lemma}
\label{le:int}
	If $u$ is a terminal for at most two out of the three commodities, then an integer stable multicommodity flow exists.
\end{lemma}

\proof
Let us now investigate the same instance with a slight modification: $s^1= s^2= t^1 = t^2 = u$, but $\{s^3, t^3\} = \emptyset$. Then, the following integer flow is stable:
$$f^1(uv_1) = f^1(v_1v_2) = f^1(v_2v_3) = f^1(v_3u) =1.$$
A blocking walk with respect to commodity~1 cannot exist, because all edges that can carry commodity~1 also carry it to their upper capacity. Commodity~2 could block along $C^2$, but edge $v_2 v_3$ is saturated with its most preferred commodity. It is trivial that the same flow remains stable if we set $s^1= t^1 = u$ and $\{s^2, t^2\} = \{s^3, t^3\} = \emptyset$. If $\{s^1, t^1\} = \{s^2, t^2\} = \{s^3, t^3\} = \emptyset$, then the empty flow is stable.
\qed

To sum up the established results about Example~\ref{ex:triangle}: the instance admits an integer stable flow if and only if $u$ has at most two commodities. This argument will help us prove a claim later in our hardness proof.

\begin{theorem}
Deciding whether \textsc{ismf} has a solution is $\NP$-complete. This holds even if all commodities share the same set of terminal vertices, all vertices have the same preferences with respect to all commodities, and edges do not have commodity-specific capacities (but edges have preferences over different commodities).
\end{theorem}

\proof
In the following, we show NP-completeness for the general version \textsc{ismf}. By Theorem~\ref{th:smf_simple}, this also implies NP-completeness for \textsc{ismf} restricted to instances with identical terminal sets, commodity-independent vertex preferences, and without commodity-specific edge capacities.

Testing whether a feasible integral multicommodity flow is stable can be done in polynomial time, as pointed out also in~\cite{KP13a}. It is sufficient to check the existence of edges fulfilling points~2 and~3 in Definition~\ref{def:mcblocking} for every commodity and then execute a breadth-first search for every pair of vertices as $v_1$ and $v_k$ vertices of the potential blocking walk. Thus \textsc{ismf} is in~$\NP$.

\begin{figure}[t]
	\centering
	\begin{minipage}{1\textwidth}
	\centering
		\begin{tikzpicture}[scale=1, transform shape,->]
		\pgfmathsetmacro{\b}{4}
		\pgfmathsetmacro{\d}{1.5}
		\pgfmathsetmacro{\a}{90}
		\pgfmathsetmacro{\c}{45}
		
		\node[vertex, label=below:$a$] (a) at (-4*\b, 0) {};
		\node[vertex, label=above:$b_1$] (b1) at (-3*\b, \d) {};
		\node[vertex, label=below:$b_2$] (b2) at (-3*\b, \d*0.6) {};
		\node[vertex, label=above:$b_{n-1}$] (bn-1) at (-3*\b, -\d*0.6) {};
		\node[vertex, label=below:$b_n$] (bn) at (-3*\b, -\d) {};
		\node[vertex, label=below:$d$] (d) at (-2*\b, 0) {};
				
		\draw [pointedline] (a) --  node[edgelabel, above, text = MyPurple, scale = 0.8, yshift=2mm] {$1, \bar{1}$} (b1);
		\draw [pointedline] (a) -- (b2);
		\draw [pointedline] (a) -- (bn-1);
		\draw [pointedline] (a) -- node[edgelabel, above, text = MyPurple, scale = 0.8, yshift=-7mm] {$n, \bar{n}$} (bn);
		\draw [pointedline] (b1) --  node[edgelabel, above, text = MyPurple, scale = 0.8, yshift=2mm] {$\bar{1}, 1$} (d);
		\draw [pointedline] (b2) -- (d);
		\draw [pointedline] (bn-1) -- (d);
		\draw [pointedline] (bn) -- node[edgelabel, above, text = MyPurple, scale = 0.8, yshift=-7mm] {$\bar{n}, n$} (d);
		
		\node[circle, scale = 0.2, fill = black] at (-3*\b, 0) {};
		\node[circle, scale = 0.2, fill = black] at (-3*\b, 0.2) {};
		\node[circle, scale = 0.2, fill = black] at (-3*\b, -0.2) {};
		\end{tikzpicture}
	\end{minipage}
	
	\begin{minipage}{1\textwidth}
	\begin{tikzpicture}[scale=0.9, transform shape,->]
	\pgfmathsetmacro{\b}{1.3}
	\pgfmathsetmacro{\d}{1.5}
	\pgfmathsetmacro{\a}{90}
	\pgfmathsetmacro{\c}{45}
	
	\node[vertex, fill=white, label=below:$s$] (s) at (-5*\b, 0) {};
	\node[vertex, label=below:$a'$] (a) at (-4*\b, 0) {};
	\node[vertex, label=above:$b_1'$] (b1) at (-3*\b, 0.5) {};
	\node[vertex] (bi) at (-3*\b, 0) {};
	\node[vertex, label=below:$b_n'$] (bn) at (-3*\b, -0.5) {};
	\node[vertex, label=below:$d'$] (d) at (-2*\b, 0) {};
	\node[vertex, label=below:$u_1$] (u1) at (-\b, 0) {};
	\node[vertex, label=below:$u_i$] (ui) at (0, 0) {};
	\node[vertex, label=below:$u_m$] (um) at (1*\b, 0) {};
	\node[vertex, label=below:$a''$] (a') at (2*\b, 0) {};
	\node[vertex, label=above:$b_1''$] (b1') at (3*\b, 0.5) {};
	\node[vertex] (bi') at (3*\b, 0) {};
	\node[vertex, label=below:$b_n''$] (bn') at (3*\b, -0.5) {};
	\node[vertex, label=below:$d''$] (d') at (4*\b, 0) {};
	\node[vertex, fill=white, label={[xshift=3mm, yshift=-7mm]$t$}] (t) at (5*\b, 0) {};
	
	\draw [pointedline] (s) -- (a);
	\draw [pointedline] (a) -- (bi);
	\draw [pointedline] (a) -- (b1);
	\draw [pointedline] (a) -- (bn);
	\draw [pointedline] (bi) -- (d);
	\draw [pointedline] (b1) -- (d);
	\draw [pointedline] (bn) -- (d);
	\draw [pointedline] (d) -- (u1);
		\node[circle, scale = 0.2, fill = black] at (-0.5*\b-0.1, 0) {};
		\node[circle, scale = 0.2, fill = black] at (-0.5*\b, 0) {};
		\node[circle, scale = 0.2, fill = black] at (-0.5*\b+0.1, 0) {};
		\node[circle, scale = 0.2, fill = black] at (0.5*\b-0.1, 0) {};
		\node[circle, scale = 0.2, fill = black] at (0.5*\b, 0) {};
		\node[circle, scale = 0.2, fill = black] at (0.5*\b+0.1, 0) {};
	\draw [pointedline] (um) -- (a');
	\draw [pointedline] (a') -- (bi');
	\draw [pointedline] (a') -- (b1');
	\draw [pointedline] (a') -- (bn');
	\draw [pointedline] (bi') -- (d');
	\draw [pointedline] (b1') -- (d');
	\draw [pointedline] (bn') -- (d');
	\draw [pointedline] (d') -- (t);
	
		\node[vertex, label=below:$v_1$] (v1) at (-\b, -2*\d) {};
		\node[vertex, label={below, yshift=-2mm}:$v_2$] (v2) at (0, -\d) {};
		\node[vertex, label=below:$v_3$] (v3) at (\b, -2*\d) {};
		
		\draw [pointedline] (ui) to[out=180+\c,in=\a] (v1);
		\draw [pointedline] (ui) to[out=180+\c,in=180-\c] (v2);
		\draw [pointedline] (ui) to[out=-\c,in=180-\a] (v3);
		
		\draw [pointedline] (v1) to[out=180-\c,in=180](ui);
		\draw [pointedline] (v2) to[out=\c,in=-\c] (ui);
		\draw [pointedline] (v3) to[out=\c,in=0]  (ui);
		
		\draw [pointedline, MyPurple] (v1) -- (v2);
		\draw [pointedline, MyPurple] (v2) -- (v3);
		\draw [pointedline, MyPurple] (v3) -- (v1);
	
	\end{tikzpicture}
	\end{minipage}
	\caption{A variable gadget and the entire construction for \textsc{ismf}.}
	\label{fi:mchardness}
\end{figure}
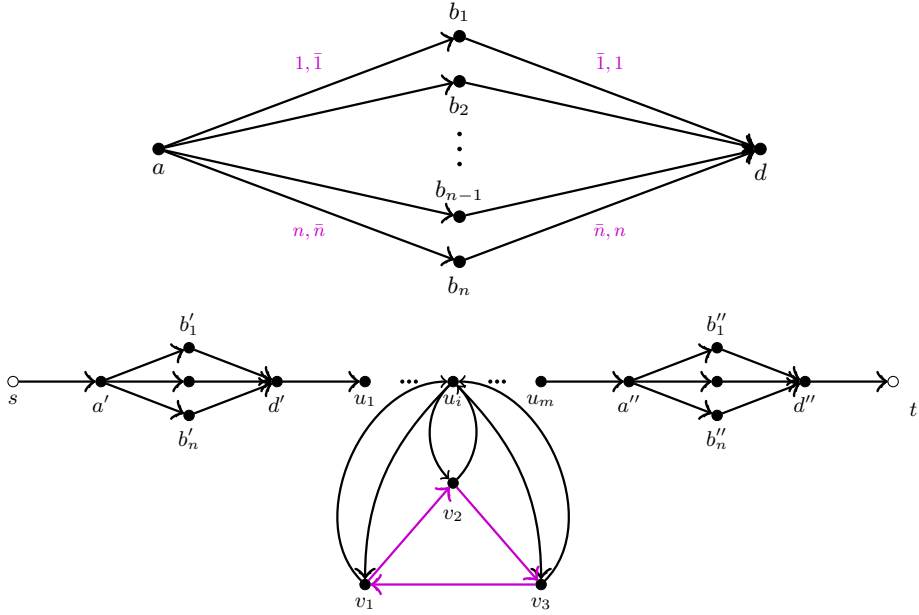

We now describe how to construct an \textsc{ismf} instance $\mathcal{I}'$ from any given instance $\mathcal{I}$ of \textsc{3-sat} with $n$ variables and $m$ clauses, also illustrated in Fig.~\ref{fi:mchardness}. For each variable $i$ in the Boolean formula we create 2 commodities, $i$ and $\bar{i}$, corresponding to truth values $\true$  and $\false$. To simplify notation, we say that $\bar{\bar{i}} = i$. Every clause in the formula is assigned a clause gadget, identical to the instance presented in Example~\ref{ex:triangle}, but with $u$ being a non-terminal for all commodities. The three \emph{relevant commodities} are the commodities corresponding to the \emph{negations} of the three literals appearing in the clause. The preferences of $u$ in such a gadget are chosen so that the edges of the gadget are preferred to edges outside of the gadget. The order of the edges at $u$ inside the gadget is irrelevant due to the commodity-specific capacity constraints.

All commodities share the same terminals $s$ and~$t$. There is a long path running from $s$ to $t$, consisting of three segments. The first and the third segments are two disjoint copies of the same variable gadget, while the second segment consists of the $u$-vertices of the $m$ clause gadgets. A variable gadget is defined on vertices $\left\{ a, b_1, b_2, ..., b_n, d \right\}$ with edges $a b_i$ and $b_i d$ for all $i$. For each~$i$ and each $e \in \{a b_i, b_i d\}$ we set the capacities $c^i(e) = c^{\bar{i}}(e) = c(e) = 1$ and $c^j(e) = c^{\bar{j}}(e) = 0$ for $j \neq i$.
Edge $a b_i$ ranks commodity $i$ best, and $\bar{i}$ second, while $b_i d$ ranks commodity $\bar{i}$ best, and $i$ second. The vertex preferences of $a$ and $d$ are arbitrary. These three segments are chained together so that the only edge of $s$ ends at $a'$ in the first variable gadget, $d'$ in the same gadget is connected to the first $u$ vertex of the second segment, the last $u$ of the same segment is adjacent to $a''$ in the second variable gadget and $d''$ in this gadget has an edge running to~$t$. For the edges connecting the segments and the $u$-vertices of clause gadgets with each other and with the terminals, the capacities are set to $c^i =  c^{\bar{i}} = c = n$ for all $1 \leq i \leq n$, and edge preferences are chosen arbitrarily.

Having described the full construction we now prove in Lemmas~\ref{lem:ismf1} and \ref{lem:ismf2} the equivalence between the existence of an integral stable multicommodity flow in $\mathcal{I}'$ and a satisfying truth assignment in~$\mathcal{I}$.

\begin{lemma}
\label{lem:ismf1}
	If an integral stable multicommodity flow $f$ exists in $\mathcal{I}'$, then there is a satisfying truth assignment in~$\mathcal{I}$.
\end{lemma}
\proof As defined after Definition~\ref{def:mf}, $f^i$ denotes the total flow value with respect to commodity~$i$.

\begin{claim}
	For every commodity $i$, $f^i + f^{\bar{i}} = 1$.
\end{claim}
	\myproof If $f^{i}(a b_i) + f^{\bar{i}}(a b_i) < 1$ for some commodity $i$ and edge $ab_i$ of a variable gadget, then there is an unsaturated $s$-$t$ path through $b_i$ with respect to commodity $i$, because the edges $a b_i$ and $b_i d$ are not saturated and all other edges along the main path have capacity~$n$. 
	This path blocks~$f$. Since $c(a b_i) = 1$ for every $1 \leq i\leq n$, $f^{i}(a b_i) + f^{\bar{i}}(a b_i) = 1$, thus edges $ab_i$ and $b_id$ of the variable gadgets are saturated with commodities $i$ and~$\bar{i}$. This already implies that $f^i + f^{\bar{i}} = 1$ for every $1 \leq i\leq n$. \myqed
	
	This claim allows us to assign exactly one truth value to each variable: $x_i$ is $\true$ if $f^i = 1$ and it is $\false$ if $f^{\bar{i}} = 1$. 
	
\begin{claim}
	For every clause $C = x_i \vee x_j \vee x_k$, where the variables in $C$ can be in negated or unnegated form, $f^{\bar{i}} + f^{\bar{j}} + f^{\bar{k}} \leq 2$, for every  $1 \leq i, j, k \leq n$.
\end{claim}
	\myproof Since $u$ prefers sending flow along its edges in the gadget over forwarding it to the next $u$ vertex on the path, $u$ can be seen as a terminal vertex with respect to the commodities reaching it. As we have shown in Example~\ref{ex:triangle}, if there is a solution to \textsc{ismf}, then at most two of the three relevant commodities are present at~$u$. \myqed
	
	The latter claim is the reason why we took the negated version of each literal in the clause: at most two literals are false in each clause, thus the clause is satisfied by the truth assignment.
\qed

\begin{lemma}
\label{lem:ismf2}
	If there is a satisfying truth assignment in $\mathcal{I}$, then there is an integral stable multicommodity flow $f$ in~$\mathcal{I}'$.
\end{lemma}

\myproof The constructed flow to the given truth assignment is the following. For every variable $i$, $f^{i} = 1, f^{\bar{i}} = 0$ if $i$ is $\true$, and $f^{i} = 0, f^{\bar{i}} = 1$ otherwise. This rule obviously determines $f$ on all edges not belonging to clause gadgets. Since we started with a valid truth assignment, each clause gadget has at most two out of the three relevant commodities $i_1, i_2$ and $i_3$ reaching~$u$. Commodity $i_j$ corresponds to commodity $j$ in Example~\ref{ex:triangle}. If one commodity $i_j$, $j \in \{ 1,2,3 \}$ is not present at $u$, then we send commodity $i_{j+1}$ (modulo~3) along cycle $C^{i_{j+1}}$ and set all other flow values in the gadget to~0. Note that this also implies that commodity $i_{j+2}$ (modulo~3) is forwarded by $u$ without entering the clause gadget. If two commodities are missing, we send the third along its cycle. If no relevant commodity reaches the gadget, then we leave all edges of the gadget empty.

We need to show now that $f$ is an integral stable flow. Feasibility and integrality clearly follow from the construction. 
Proceeding from $s$ to $t$ in the graph, we investigate at which vertex a blocking walk $W$ might start. 

\begin{enumerate}
	\item Assume $W$ starts at $s$. If $a' b'_j$ is the edge saturated by its best commodity, then $W$ cannot proceed through $a' b'_j$. If $a' b'_j$ is not saturated by its preferred commodity, then $b'_j d'$ is and $W$ cannot pass through $b'_j d'$.
	Hence $W$ either ends at $a'$ or $b'_j$ for some $j$. In either case, it ends at a non-terminal vertex with a single incoming edge. Thus a walk $W$ starting at $s$ cannot block $f$.
	\item Similarly, if $W$ starts at $a'$, it has to end at $b'_j$ for some $j$ and thus $W$ cannot block $f$.
	\item For each $j$, the non-terminal vertex $b'_j$ has a single outgoing edge. Thus it also cannot start a blocking walk.
	\item The same holds for~$d'$.
	\item The same arguments apply for walks starting at $a''$, $b''_j$ for some $j$, or $d''$, respectively.


	\item If $W$ starts at a vertex $u_j$, then its first edge must be in a clause gadget, because the edge running outside of the clause gadget is the least preferred outgoing edge of~$u_j$.
	
	Assume now without loss of generality that the first edge of $W$ is $u_j v_1$ in some clause gadget with relevant commodities $i_1, i_2$ and $i_3$, in this order. Because $u_j v_1$ only admits flow of commodity $i_1$, the walk $W$ can only be blocking with respect to commodity $i_1$, and $f^{i_1}(e) = 1$ on the edge $e$ leaving $u_j$ outside the clause gadget.
	Thus, $u_j v_1$ is not saturated, which means that commodity~$i_1$ was not chosen to fill~$C^1$. According to our rules above, the only reason for this is that commodity~$i_2$ is not present at $u$ and commodity~$i_3$ saturates~$C^3$. Then the only edge that could be the second edge of $W$ is $v_1 v_2$ in the gadget, but this edge is saturated by its best ranked commodity~$i_3$. We conclude that a blocking walk cannot start at $u_j$ for any $j$.
	
    \item Now assume $W$ starts at a vertex $v$ in the interior of a clause gadget attached to $u_j$. Without loss of generality, let this vertex be $v_1$. 
    Note that $v_1$ has two outgoing edges $v_1 v_2$ and $v_1 u_j$, but $v_1 v_2$ only supports flow of commodities $i_1$ and $i_3$, whereas $v_1 u_j$ only supports flow of commodity $i_2$. A walk starting with $v_1 u_j$ can only block $f$ with respect to commodity $i_2$, but then it cannot dominate $f$ at the start because $f^{i_2}(v_1 v_2) = 0$. Likewise, a walk starting with $v_1 v_2$ can only block $f$ with respect to $i_1$ or $i_3$, but cannot dominate $f$ at the start because $f^{i_1}(v_1 u_j) = f^{i_3}(v_1 u_j) = 0$.
	
	\item No edge leaves~$t$, so $W$ cannot start with $t$.
\end{enumerate}
We thus eliminated all possible starting vertices for blocking walks.
Since no walk blocks the constructed flow, it is stable. \myqed

\section{Conclusion and open problems}

In this paper we presented four results:
\begin{enumerate}
	\item a polynomial version of the Gale-Shapley algorithm for stable flows;
	\item a direct algorithm for stable flows with restricted intervals;
  \item a simplification of the stable multicommodity flow problem;
	\item the $\NP$-completeness of the integral stable multicommodity flow problem.
\end{enumerate}

A natural open question regarding the problem of stable flows with restricted edges presented in Section~\ref{sec:re_sf} is that of approximation.
The approximation concept of minimum number of blocking edges or minimum number of violated restrictions~\cite{CM16} can be translated to \textsc{sf restricted}. Even if there is no stable flow saturating all forced edges or avoiding all forbidden edges, how can stability be relaxed such that all edge conditions are fulfilled? Or the other way round: how many edge conditions must be violated by stable flows? 

The big open question of Section~\ref{sec:smcf} is clearly algorithms for finding a (possibly fractional) stable multicommodity flow. Even though Theorem~\ref{th:ppad} states that it is $\PPAD$-hard to find a solution in the general case, it is natural to ask whether this complexity changes when restricting the number of commodities, the maximum degree, or other parameters of the instance. Since the Gale-Shapley algorithm typically executes steps with integer values if the input is integral and we showed the hardness of \textsc{ismf}, it is likely that a novel approach is needed. Linear programming is a promising direction, but constructing a description of the \textsc{smf} polytope seems to be an extremely challenging task. At the moment, the most elaborate structure for which a linear program is known is many-to-many stable matchings~\cite{Fle03a}.

Finally, all stable flow models discussed in this paper can be combined with other common notions in stability or flows, such as ties in preference lists, edge weights, unsplittable flows, and so on.

\paragraph{Acknowledgment} We thank Tam\'as Fleiner for discussions on Lemma~\ref{le:singleforced}, and our reviewers for their suggestions that significantly improved the presentation of the paper.
\bibliographystyle{spmpsci}
\bibliography{mybib}
\end{document}